\documentclass[]{sn-jnl}

\usepackage{graphicx}  % Ensure graphics package is loaded
\usepackage{amsmath}   % Math support
\usepackage{hyperref}  % Hyperlinks (optional)
\usepackage[table,svgnames]{xcolor}
\usepackage{setspace}

\usepackage{subcaption} 
\usepackage[export]{adjustbox}
\usepackage{geometry}

\usepackage{longtable}
\usepackage{booktabs} % For professional-looking rules (\toprule, \midrule, \bottomrule)
\usepackage{array}   % For advanced column formatting

\usepackage{listings}

\usepackage{natbib}

\usepackage{xcolor}
\definecolor{codegreen}{rgb}{0,0.6,0}
\definecolor{codegray}{rgb}{0.5,0.5,0.5}
\definecolor{codepurple}{rgb}{0.58,0,0.82}
\definecolor{backcolour}{rgb}{0.95,0.95,0.92}
\definecolor{keywordblue}{rgb}{0.13, 0.13, 1}
\definecolor{variableorange}{rgb}{1, 0.5, 0}

\newcommand\listingfontsize{\fontsize{6pt}{8pt}\selectfont}

\lstdefinestyle{sparqlstyle}{
   backgroundcolor=\color{backcolour},  
   commentstyle=\color{codegreen},
   keywordstyle=\color{keywordblue}\bfseries,
   stringstyle=\color{codepurple},
   numberstyle=\tiny\color{codegray},
   basicstyle=\ttfamily\listingfontsize,
   breakatwhitespace=false,      
   breaklines=true,           
   captionpos=b,             
   keepspaces=true,           
   numbers=left,             
   numbersep=5pt,            
   showspaces=false,           
   showstringspaces=false,
   showtabs=false,            
   tabsize=2,
   morekeywords={SELECT, WHERE, PREFIX, DISTINCT, rdf, ai4se, type, hasLevel, hasTarget},
   comment=[l]{\#}, % Comments start with #
   moredelim=**[is][\color{variableorange}]{?}{ }, % Style for variables like ?paper
}

\title{Augmenting software engineering with AI}
\subtitle{The \textit{ai4se} taxonomy and its use}

\author*{Ina K. Schieferdecker}
\affil*{Technische Universit{\"a}t Berlin, Einsteinufer 25, 10587 Berlin, Germany}
\email{ina.schieferdecker@tu-berlin.de}

\abstract{
	Although model-driven software engineering (MDSE) has proven effective in managing complex systems, its industrial adoption remains limited by the substantial maintenance overhead required for models and the specialised skills demanded of developers. Meanwhile, advances in artificial intelligence (AI), particularly generative and agentic AI, have shown great promise in automating code-related tasks such as comprehension, generation, and defect detection. These capabilities are largely powered by 'big code': vast repositories of open-source software that now form the basis of data-driven, empirical SE and automated quality assurance. This paper aims to synthesise these two domains by exploring the integration of AI into model-driven practices. It provides a comprehensive overview of the current state of AI-augmented software engineering and introduces a novel taxonomy 'ai4se' to classify and connect diverse AI applications within the field. On this basis, the paper proposes a vision for 'big models' in software engineering (SE), an approach designed to leverage the structural advantages of MDSE alongside the scalability of AI. Finally, the paper discusses the pair modelling paradigm as a collaborative framework for the MDSE industry, designed to enhance software quality through human–AI partnership.}

\begin{document}

%\the\linewidth

%%%%%%%%%%%%%%%%
\maketitle         

%%%%%%%%%%%%%%%%
\clearpage
\section{Introduction}

\begin{flushright}
\begin{minipage}[t]{0.55\textwidth} 
\footnotesize 
\begin{spacing}{1} 
"In 12~months, we may be in a world where AI is writing essentially all of the code." \\Dario Amodei, CEO Anthropic, Mar. 10, 2025
\end{spacing}
\end{minipage}
\end{flushright}

The rapid advancement of generative artificial intelligence (AI) has triggered a fundamental shift in software engineering (SE). As AI tools evolve from basic code assistants into autonomous agents, the SE discipline is undergoing a transformation that will change the way software is designed, developed and maintained. However, this transition remains fragmented. Although AI research is progressing rapidly, there is currently no systemic framework in place to classify these advancements and connect them to established SE paradigms.

This paper bridges this gap by introducing the \textit{ai4se} taxonomy. This novel framework classifies AI applications in SE across four core dimensions: the purpose of the AI usage, the targeted SE activity (e.g., requirements or testing), the type of AI utilised (e.g., subsymbolic, generative, or agentic), and the level of autonomy granted to the AI.

The aim of SE is to provide systematic approaches for building high-quality, reliable and maintainable systems. However, even well-established SE practices often find it difficult to manage the inherent complexity of modern software. High-profile system outages, such as the 2024 CrowdStrike incident, underscore the challenges of validating large-scale, complex systems~\citep{crowdstrike2024}. Such incidents reflect the fact that academic breakthroughs in SE, such as Model-Driven Software Engineering (MDSE)~\citep{selic2003pragmatics,schieferdecker2024}, often encounter a significant delay in being adopted in everyday industrial practice. It is clear that the gap between rigorous, model-based research and practical, code-centric development remains a significant challenge.

This paper explores whether AI can serve as the catalyst to bridge this divide and investigates specifically how AI can support MDSE to improve software quality. As \citet{brooks1987essence} noted, the greatest challenges in software development including specification, design, and testing are "essential" to the complexity of the system, rather than the "accidental" work of writing code. While past high-level languages addressed accidental difficulties, MDSE aims to tackle essential complexity through abstraction. Integrating AI into MDSE could potentially lower the barrier to adopting these disciplined, model-centric practices.

This intersection is investigated through a Systematic Literature Review (SLR) and a rigorous classification of recent publications. The paper’s main contributions are:
\begin{itemize}
	\item The \textbf{\textit{ai4se} taxonomy}: A framework to categorise the purpose, target, type, and autonomy of AI for SE research.
	\item The \textbf{Model Naturalness Hypothesis}: A proposition that model corpora contain statistical properties exploitable for AI tool development for MDSE.
	\item The \textbf{Big Models concept}: 	A concept analogous to 'big code' in SE treating models as large-scale datasets, which can be engineered top-down or extracted bottom-up from existing codebases, enabling the application of sophisticated AI-driven analysis.
	\item The \textbf{Pair Modelling paradigm}: A collaborative model where AI agents and software engineers co-evolve models and code, fostering software quality and ethical oversight\footnote{Key ethical dimensions in MDSE encompass transparency and understandability, bias and fairness, accountability and responsibility, security and privacy, control and autonomy, intellectual property and licensing, and environmental impact. While these ethical considerations are pertinent to any human-made, technical, or software-based system, MDSE introduces distinct challenges due to its inherent reliance on abstraction. The transformation of high-level models into executable code can obscure design decisions, embed subtle biases, diffuse responsibility, and create opaque system behaviours in ways that differ significantly from traditional code-centric development. This unique interplay of abstraction and ethics within MDSE warrants dedicated exploration and merits a separate, in-depth publication to fully elaborate its implications and propose effective mitigation strategies.}.
\end{itemize}

The remainder of this paper is structured as follows: Section~\ref{Back} provides background information on big code in SE, on MDSE and on AI in SE. Section~\ref{AI4SE} presents the taxonomy \textit{ai4se}, providing details on the systematic literature review process and the classification of recent research on the application of AI for SE. Section~\ref{Related} discusses related work. Section~\ref{Sum} concludes the paper with a summary and outlook.

%%%%%%%%%%%%%%%%
\section{Background}\label{Back}

%%%%%%%%%%%%%%%%
\subsection{Big Code in Software Engineering}\label{Big}

The online availability of open-source software (OSS) paved the way for software reuse. It has also transformed the landscape of software licences and the way software is developed, with developers using OSS for specific components or entire products. OSS has also enabled deeper insights into software architecture and development practices. It has enabled new approaches to empirical software engineering, deepening our understanding of SE.

\begin{figure}[!ht]
	\centering % This centers the entire figure block
	
	\begin{subfigure}{\textwidth}
		\centering
		% The key is to set the width relative to the subfigure's container
		\includegraphics[width=0.7\linewidth]{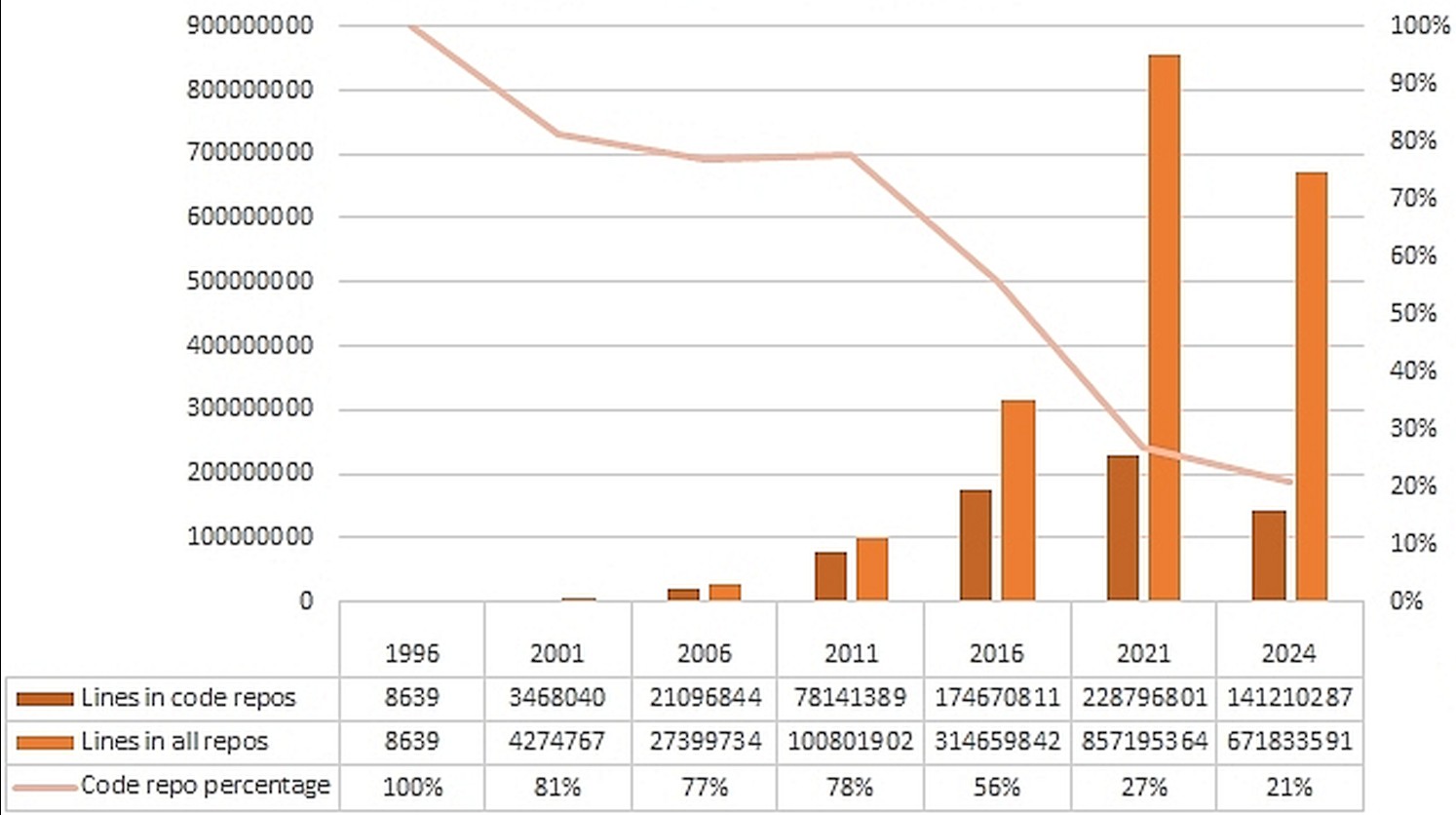} 
		\caption{Lines and portion of GitHub code repositories}
		\label{figAllRepos}
	\end{subfigure}
	
	\begin{subfigure}{1\textwidth}
		\centering
		% The key is to set the width relative to the subfigure's container
		\includegraphics[width=0.7\linewidth]{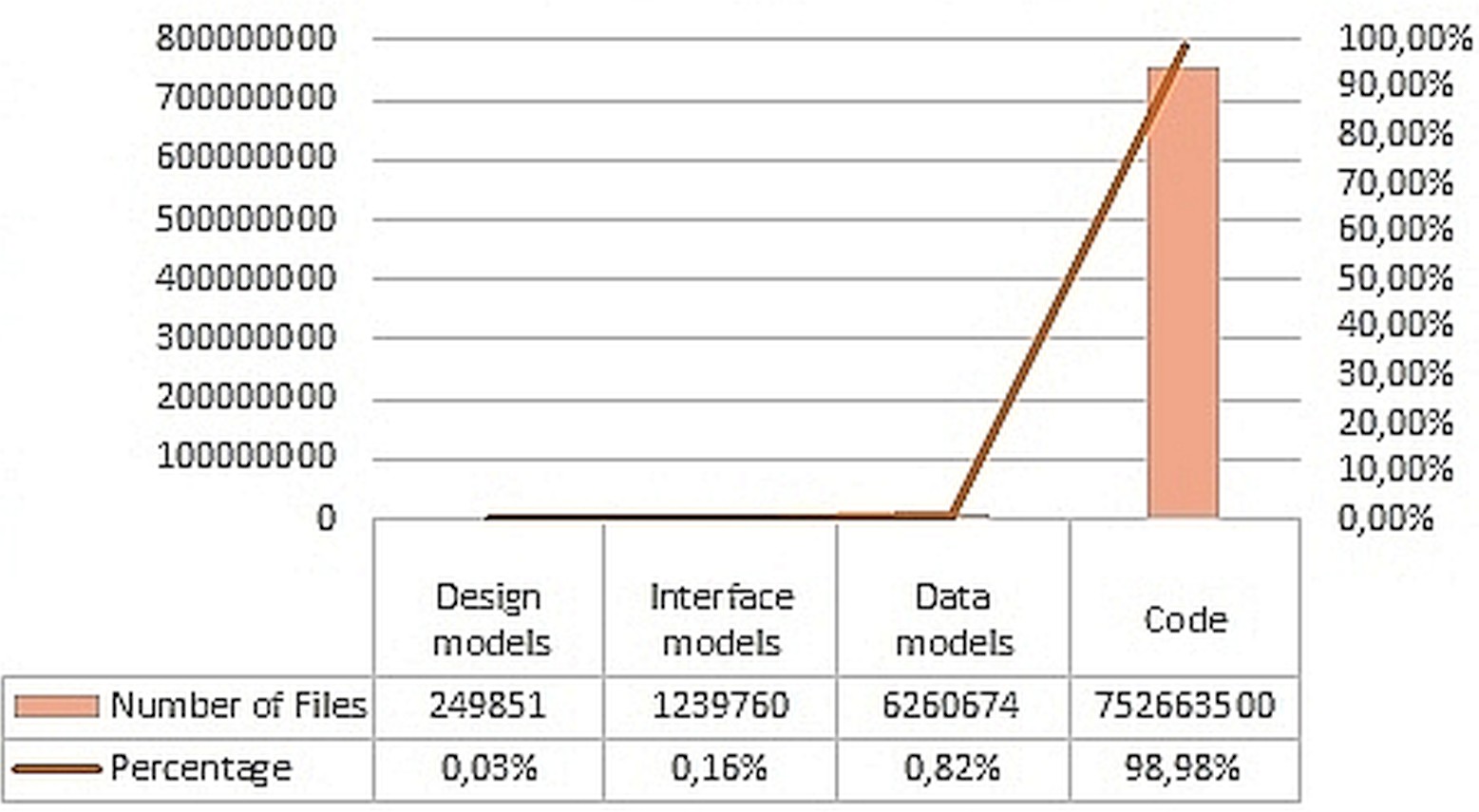} 
		\caption{Number and percentage of modelling and coding files in GitHub repositories}
		\label{figAllFiles}
	\end{subfigure}
	
	\begin{subfigure}{0.9\textwidth}
		\centering
		\includegraphics[width=0.7\linewidth]{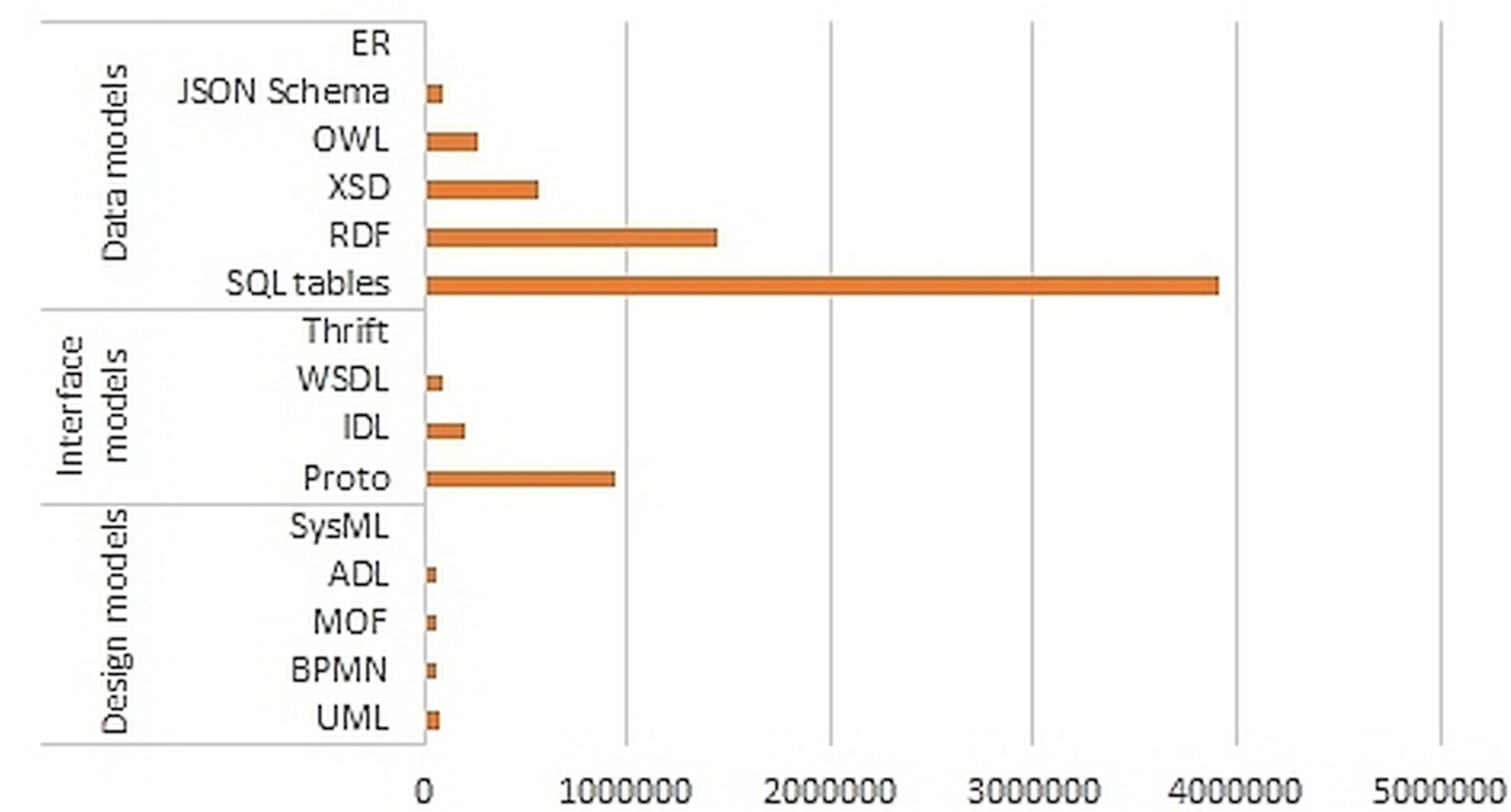}
		\caption{Number of modelling files in GitHub repositories}
		\label{figModelFiles}
	\end{subfigure}
	
	\caption{Repositories breakdown by Kibble and GitHub search taken 4-6 February 2025}\label{figGitHub}
\end{figure}

The advent of software coding platforms can be traced back to the 1990s, although it was not until the early 2000s that they truly gained prominence, with the launch of GitHub in April 2008 representing a significant milestone in this regard. At the present time, GitHub is the most widely utilised software coding platform. As stated by~\citet{zee2021}, GitHub hosts over 100~million projects and 40~million users. It became not only a significant platform for software engineering collaboration, but also a prominent reference for open-source software mining~\citep{kalliamvakou2014promises}. The study also demonstrated that a considerable number of GitHub repositories are not directly related to software development. This is because GitHub is not solely utilised for coding collaborations; it is also employed for collaborations on websites, editing books or other publications, and is even used as a storage platform. A later manual analysis by the same authors~\citep{kalliamvakou2016depth} revealed that over a third of the repositories on GitHub were not software development repositories. According to~\citet{kibble2025}, over 20\% of the total lines stored on GitHub reside in code repositories (see Figure~\ref{figAllRepos}). By utilizing GitHub's ability to search across all repositories by programming or modelling language, these files were categorized into design models, interface models, data models, and source code as illustrated in Figure~\ref{figAllFiles} and Figure~\ref{figModelFiles} and discussed in Section~\ref{Models}.

As the vast quantities of software code have accumulated over time, the term "Big Code" has evolved  (see, for example,~\citep{markovtsev2018public,ortin2016big,vechev2016programming}): Not only can the available code be reused, extended and analysed, it can also be used to train ML models, enabling the application of AI for SE on a large scale~\citep{li2023starcoder}. The term big code signals a fundamental shift in the analysis and construction of software, moving from a focus on the formal logic of a single program to identifying statistical patterns within vast repositories of source code. \citet{allamanis2014learning} considered the statistical properties of code to be similar to those of natural language, meaning code is full of learnable, predictable patterns. This observation later led to the formulation of the naturalness hypothesis~\citep{allamanis_survey_2017}. 

%%%%%%%%%%%%%%%%
\subsection{Models in Software Engineering}\label{Models}

\begin{figure}
	{\centering \includegraphics[scale=0.22]{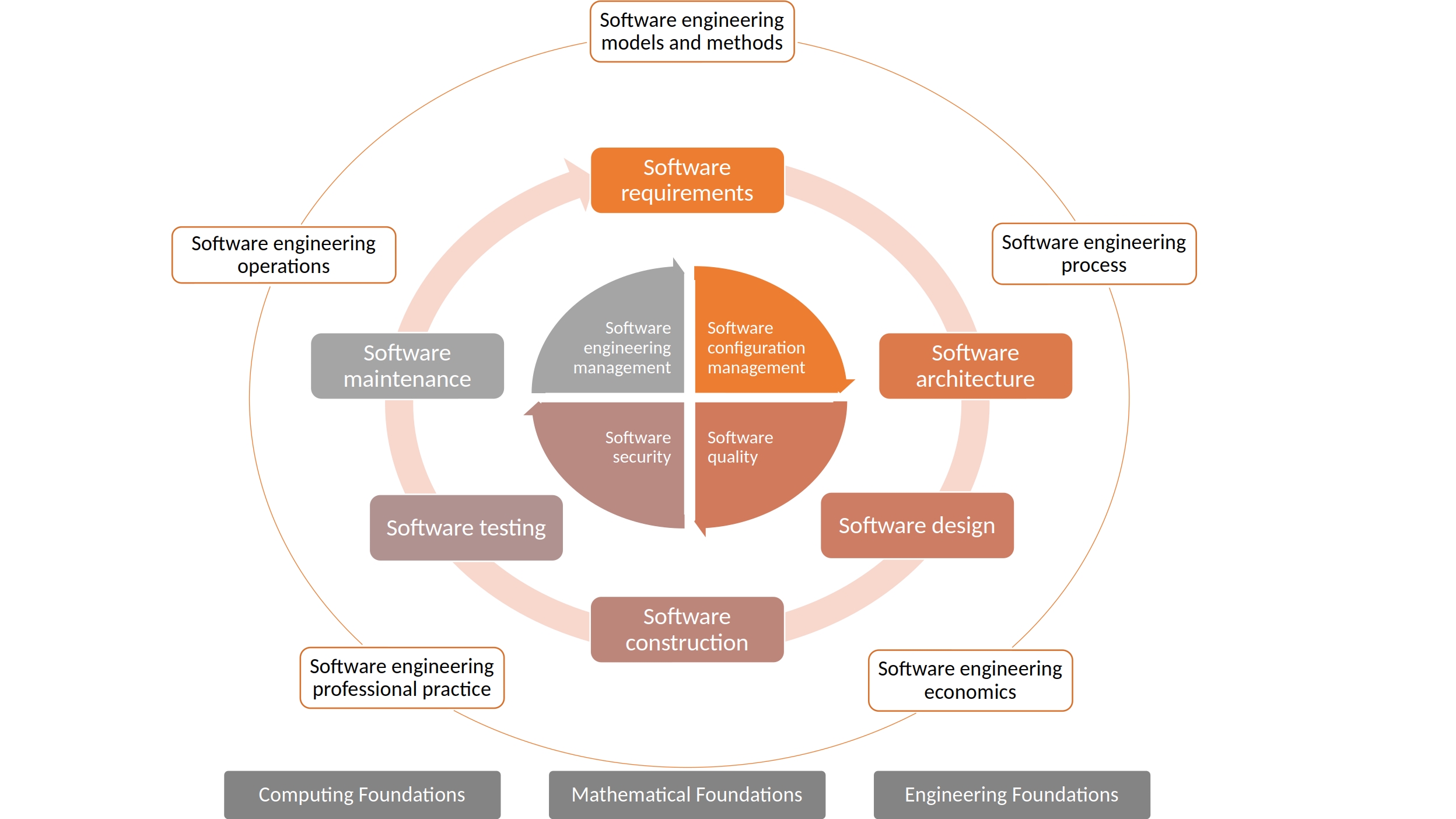}\par}
	\caption{Software Engineering knowledge areas according to SWEBOK~\citep{swebok2024}} \label{figSWEBOK}
	\end{figure}

According to the IEEE Software Engineering Body of Knowledge (SWEBOK~\citep{swebok2024}, see also Figure~\ref{figSWEBOK}), the software lifecycle is comprised of distinct phases and activities that can be described as knowledge areas, including requirements, architecture, design, construction, testing, and maintenance. There are also knowledge areas that deal with the fundamentals of computer science, mathematics, and engineering, as well as cross-cutting activities pertaining to software quality, security, configuration management, and engineering management. Furthermore, there are cross-cutting knowledge areas that encompass the processes, operations, professional practice, and economics of software engineering, along with a distinct area for software engineering models and methods. This knowledge area encompasses the processes and methodologies associated with modelling, the various types of models, including those pertaining to information, behaviour, and structure, and the analysis of models, see also~\citep{schieferdecker2024}. Overall, SWEBOK addresses dedicated model-based and model-driven approaches in each knowledge area, such as model-based requirements specification, model-based architecture, model-driven design, lifecycle models, executable models, model-based testing, etc.

This is because models in SE are the key to dealing with the complexity of software: Models provide the essential abstractions to capture requirements, support design decisions, or to offer comprehensive overviews of software structures and behaviours. They are essential tools in the engineering process for constructing and maintaining software. They are also vital for configuration, monitoring, and runtime support during software operation and management. 

However, the adoption of MDSE in the industrial practice remains limited. \citet{alfraihi2023trends} reports that, in this survey, only 37.4\% of the participants stated that "MD[S]E is used in most projects and by most developers", while 46\% reported limited or no use. Nowadays, however, MDSE does not only rely on models in the Unified Modelling Language (UML), in other standardized modelling languages, such as Business Process Modelling Language (BPMN), or in Domain-Specific Languages (DSL): Rather than explicitly developing software models and using them from the top down, they can be extracted from software executions, monitoring, or static analysis. These approaches aim to minimise the costs associated with model development while still reaping the benefits of MDSE~\citep{bagheri_bottom-up_2013}. The combination of explicitly developing model artefacts and extracting models from code artefacts has resulted in a hybrid top-down/bottom-up MDSE approach, as described in \citep{steffen2007model}. This approach benefits from both directions of MDSE, as demonstrated in~\citep{vaupel2015agile,weissleder2013top,garcia2023lifting}. Furthermore, as large volumes of data are interpreted more dynamically depending on their current application context, new data modelling approaches have evolved~\citep{ribeiro2015data}. Examples include Not Only SQL (NoSQL), JavaScript Object Notation (JSON), Resource Description Framework (RDF), and Yet Another Markup Language (YAML). 

The formal structures of models, together with the increased use of advanced data models and the bottom-up and hybrid modelling approaches, have paved the way for MDSE with AI by making larger model sets available for AI training. Additionally, machine learning is employed to render models suitable for further processing, even when they are not inherently machine-processable, as with images:~\citet{shcherban_multiclass_2021} presents an approach to automatically detecting images containing one or more of the ten types of UML diagrams being analysed. This approach was used to derive a larger model set containing these ten types of UML diagrams. It complements other approaches for providing ground truth model sets, such as the Lindholmen model set~\citep{robles_extensive_2017,robles_reflection_2023},  which includes 'almost 100k models from 22k [GitHub] projects',  and the ModelSet~\citep{lopez_modelset_2022}.

An analysis using the GitHub search API indicates that around 1\% of GitHub files contain models or partial model information. Counting the number of code and model files in GitHub repositories reveals the ratio of code to model files for software designs, interface models, and data models (see Figure~\ref{figAllFiles}). Figure~\ref{figModelFiles} shows the split between design, interface and data models. These figures are only rough indicators. Lines of code and the number of files are crude measures of software. File extensions do not accurately indicate file types. Furthermore, there are numerous coding and modelling tools, some of which have proprietary file extensions. Images are not counted.

This analysis showing a low number of modelling files on GitHub aligns with~\citet{france2013so}, who noted over a decade ago that the high overhead of round-trip engineering often outweighed the benefits of MDSE. These historical limitations explain why the recent GitHub analysis reveals that design and behaviour models are still rarely used today. Nevertheless, both the Lindholmen dataset and this GitHub analysis demonstrate that modelling has not been abandoned. Rather, it is predominantly employed for data and interface structures, as their robust functionality heavily depends on well-defined schemas~\citep{crowdstrike2024}. While emerging AI tools hold the potential to resolve past MDSE limitations by automating model synchronization, current repository data indicates that developers continue to favor modelling for data and interfaces rather than overall software design.

%%%%%%%%%%%%%%%%
\subsection{Artificial Intelligence in Software Engineering}\label{AI}

With Big Code (see Section~\ref{Big}) serving as the crucial training ground, AI models have gained unprecedented ability to recognize coding patterns and anti-patterns. This advancement has made it possible for a suite of sophisticated AI-driven tools to emerge, offering capabilities such as code completion, defect detection, code summarization, translation, enhanced code searching, and the checking and learning of coding conventions, among other SE functions.

The potential of using AI for SE was discussed as early as 1988 in~\citet{barstow_artificial_1988}, and has been a topic of discussion ever since, most recently for instance in~\citet{alenezi_ai-driven_2025}. For example,~\citet{feldt_ways_2018} proposes the AI in SE Application Levels (AI-SEAL) taxonomy, which differentiates between the point of applying AI to the software engineering process, the software product or at runtime, the levels of automation, and the types of AI along the five tribes differentiation by~\citet{domingos2015master}. It is noteworthy that the majority of the papers analysed in~\citet{feldt_ways_2018} employ AI in the software engineering process. However, despite the numerous facets of the software engineering process (see Figure~\ref{figSWEBOK}), this study does not provide further elaboration. Nevertheless, other publications offer more detailed discussions of AI in SE, including~\citet{barenkamp_applications_2020} or~\citet{ozkaya_application_2023}, see also Section~\ref{Related}. Notwithstanding the aforementioned considerations, a taxonomy for the role of AI in software engineering has yet to emerge.

In light of the recent advancements in machine learning (ML), in particular in deep learning (DL), that have made significant breakthroughs possible, this paper focuses mainly on the latest developments in the application of ML to SE. For ML to be effective, it is essential to utilise the appropriate structures inherent to SE artefacts, which has been a topic of considerable debate:~\citet{allamanis_survey_2017} presents an overview of the various ways in which source code can be represented, including representational models of tokens, token contexts, program dependency graphs, API calls, abstract syntax trees, object usage, and others. These representational models are used for AI assistance in SE, including the creation of recommender systems, the inference of coding conventions, the detection of anomalies and defects, the analysis of code, the rewriting and translation of code, the conversion of code to text for the purposes of documentation and information retrieval, and the synthesis and general generation of code from text. \citet{karampatsis_big_2020} adds further applications of AI assistance such as code completion, API migration and code repair. Furthermore,~\citet{allamanis_survey_2017} presents "\textbf{[t]he naturalness hypothesis}[:] Software is a form of human communication; software corpora have similar statistical properties to natural language corpora; and these properties can be exploited to build better software engineering tools.". Given the intrinsic formats and formal characteristics of coding and modelling languages employed in SE, the second aspect of the naturalness hypothesis can be considered relatively straightforward. Yet, the initial proposition reinforces the necessity for SE artefacts that are readily comprehensible. This assertion was previously made by~\citet{fowler2005refactoring} in a different form: "[A]ny fool can write code that a computer can understand, good programmers write code that humans can understand.". Consequently, the application of AI in SE is not merely concerned with code generation or code repair; it also encompasses the enhancement of code through techniques such as refactoring, with the objective of optimising readability and maintainability. 

The scientific literature reveals a growing landscape of approaches leveraging AI in SE. These can be broadly categorized into three principal lines of inquiry: enhancing the understanding of SE artefacts, generating new SE artefacts from existing ones, and optimizing their quality and efficiency. These core approaches are applicable across the entire software development lifecycle, impacting diverse artefacts such as requirements, designs, code, tests, and build configurations, as well as various SE processes. To systematically capture the multifaceted aspects, perspectives, and properties of integrating AI for advancing software engineering methods and processes, the \textit{ai4se} taxonomy has been developed.

%%%%%%%%%%%%%%%%
\section{The taxomony \textit{ai4se} on Artificial Intelligence in Software Engineering}\label{AI4SE}

\begin{figure}[!ht]
	{\centering \includegraphics[width=\linewidth, 
		max height=0.55\textheight]{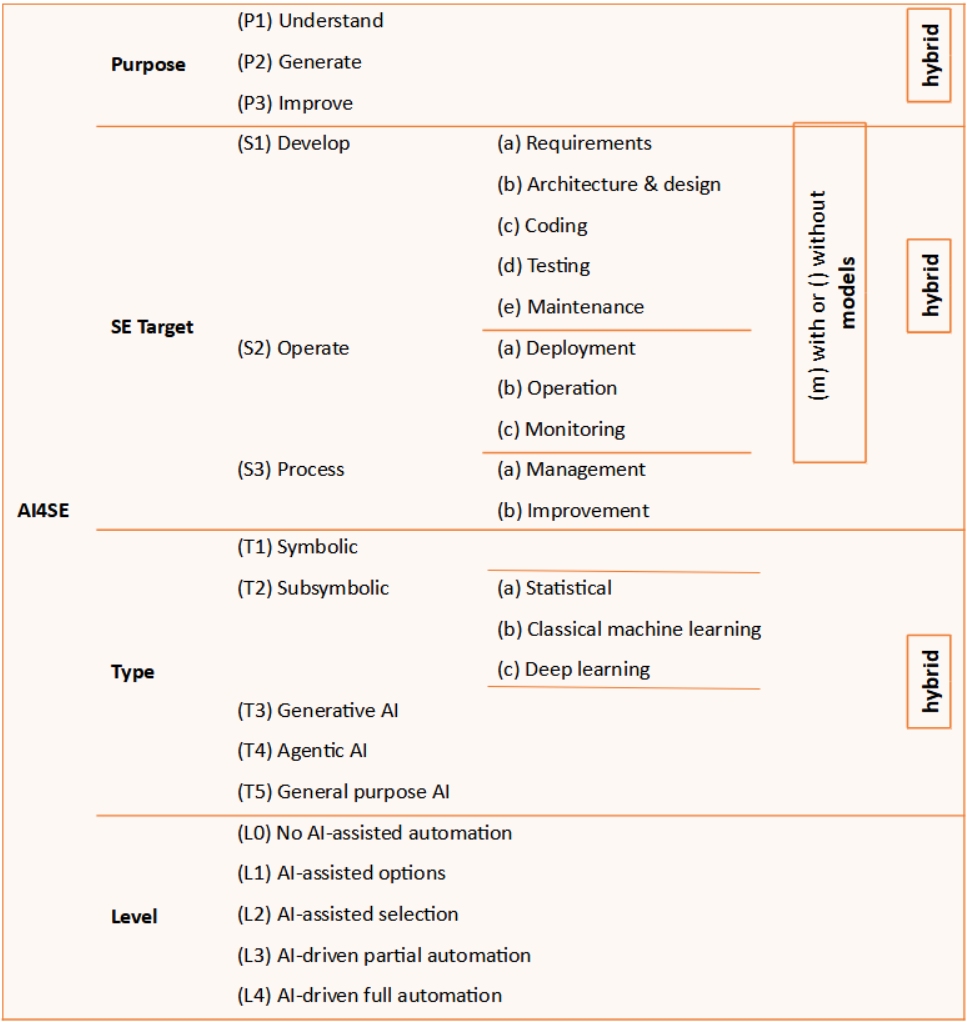}\par}
	\caption{Overview of the \textit{ai4se} taxonomy} \label{figAI4SE}
\end{figure}

The \textit{ai4se} taxonomy (see Figure~\ref{figAI4SE}) was inspired by numerous discussions concerning the application of AI in the field of SE. In the absence of a taxonomy that addressed the various options and intricacies of SE and/or AI, a new taxonomy was developed, drawing inspiration from other surveys (see Section~\ref{Related}). The dimensions of the first level of \textit{ai4se} are categorised as follows: (1)~the \textbf{purpose} of using AI for software engineering, (2)~the \textbf{target} of using AI within software engineering, (3)~the \textbf{type} of AI used, and (4)~the \textbf{level} of autonomy/automation achieved by the AI.

The \textbf{purpose} dimension has the three facets of (P1)~\textbf{understanding} to gain insight, (P2)~\textbf{generating} to derive (parts of), and (P3)~\textbf{improving} to refine SE artefacts and/or SE processing. An approach may address multiple purposes. If it is \textbf{hybrid} in this sense, all its purposes are made explicit, e.g. to support understanding only is denoted by 'P1', or to support both understanding and generation is denoted by 'P1, P2'. 

The \textit{ai4se} application \textbf{targets} consist of the main SE activities as described by SWEBOK~\citep{swebok2024} and DevOps~\citep{kumar_assessment_2024}: (S1)~Software \textbf{development} includes requirements engineering, architecture and design, coding, testing, and maintenance. (S2)~Software \textbf{operations} includes deployment, operation, and monitoring. Together with the (S3)~SE \textbf{process} including management and improvement, the three form the target of application dimension in the taxonomy. Each of the target options can be performed with the support of models, based on models or even driven by models, or without them. For the sake of comprehension of the taxonomy, the \textbf{model} option is optional for each application target: If models are used, i.e. the application target is MDSE, it is marked with "(m)". Thus, the application target "S1.m" refers to software development with models, and "S1" to one without. Or, in a more fine-grained analysis, "S1.d.m" would be used for model-based testing and "S1.d" for testing without models. An approach may address several application targets. If it is \textbf{hybrid} in this sense, all of its goals can be made explicit, e.g. support for coding only is denoted by "S1.c", support for both coding and testing is denoted by "S1.c, S1.d", or support for coding and model-based testing is denoted by "S1.c, S1.d.m". For simplicity, "h" can also be used to indicate a hybrid approach of several targets. 

The \textbf{types} of AI are divided into (T1)~\textbf{symbolic} AI, i.e. knowledge representation, rule-based systems, and logical reasoning, (T2)~being \textbf{subsymbolic} including \textbf{statistical} AI, i.e. probabilistic modelling and inference, Bayesian networks, Markov models, Gaussian processes, etc., \textbf{classical machine learning}, i.e. supervised, unsupervised, and reinforcement learning, feature engineering and selection, etc., and \textbf{deep learning}, i.e. neural networks, transformers, transfer learning, self-supervised learning, attention mechanisms, etc., (T3)~\textbf{generative AI} (GenAI), i.e. large language models (LLM), generative adversarial networks, diffusion models, variational autoencoders, autoregressive models, contrastive learning, reinforcement learning from human feedback, etc., (T4)~\textbf{agentic AI}, i.e. multi-agent systems, planning and decision making, hierarchical reinforcement learning, etc., and (T5)~\textbf{general purpose AI}, i.e. meta-learning and world models. Today, all types (T1) to (T5) are used in software engineering, but not (T5) due to its non-existence. An approach can also use a combination of AI types, i.e. it can be \textbf{hybrid}. This can be made explicit, e.g. by denoting "T1, T2" for a hybrid approach that uses both symbolic and classical machine learning. For simplicity, "h" can just be used to denote a hybrid approach that uses more than one type of AI. 

The \textbf{levels} of autonomy/automation are defined similarly to the levels of autonomous driving~\citep{barabas2017current} as five levels from no support to recommendations to full automation: (L0)~\textbf{No} AI-assisted automation, (L1)~AI-assisted \textbf{options}, where the developer is in full control and receives recommendations to choose from and adapt, (L2)~AI-assisted \textbf{selection}, where the AI selects pre-defined options, (L3)~AI-based \textbf{partial automation}, where the AI selects options in simple, standard cases, and (L4)~AI-based \textbf{full automation}, where the AI operates without the developer. So far, level L1 and L2 are the most common, level L3 is on the rise and level L4 is rather unrealistic for complex, industrial-scale software. Higher levels of autonomy typically include the capabilities of lower levels of autonomy; when presenting the level of autonomy of an approach, the highest level of autonomy is given.

\begin{figure}[!ht]
   \centering % This centers the entire figure block

   \begin{subfigure}{0.54\textwidth}
     % The key is to set the width relative to the subfigure's container
     \centering
     \includegraphics[width=\linewidth, 
        max height=0.23\textheight]{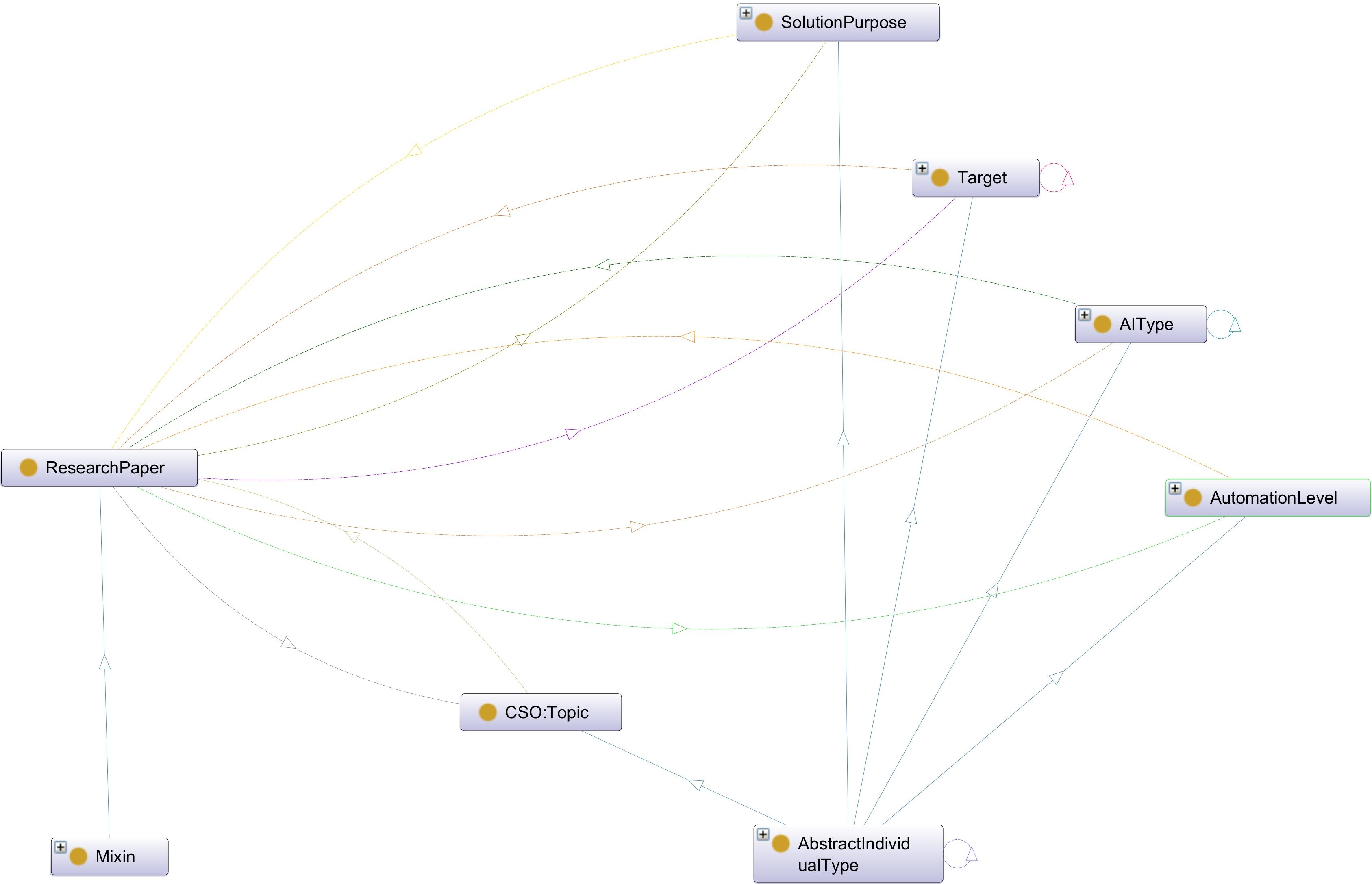} 
     \caption{Overview of the \textit{ai4se} ontology}
     \label{figai4seontology}
   \end{subfigure}% <--- This % sign is important, it prevents a horizontal space
\hfill
   \begin{subfigure}{0.44\textwidth}
     \centering
     \includegraphics[width=\linewidth, 
        max height=0.15\textheight]{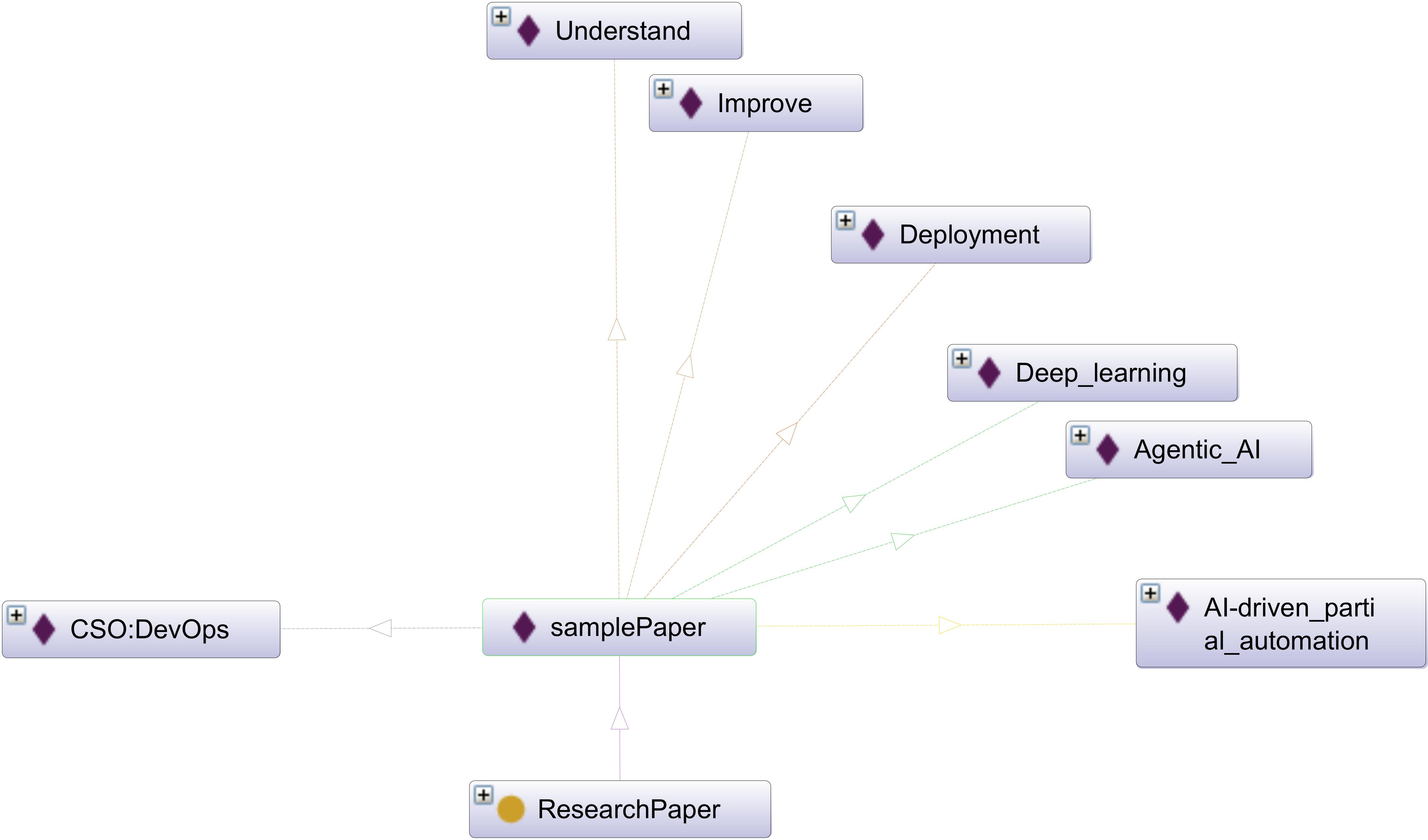}
     \caption{Sample paper classified in \textit{ai4se}}
     \label{figai4sesample}
   \end{subfigure}

   \begin{subfigure}{0.9\textwidth}
     \centering
     % The key is to set the width relative to the subfigure's container
     \includegraphics[width=\linewidth, 
        max height=0.29\textheight]{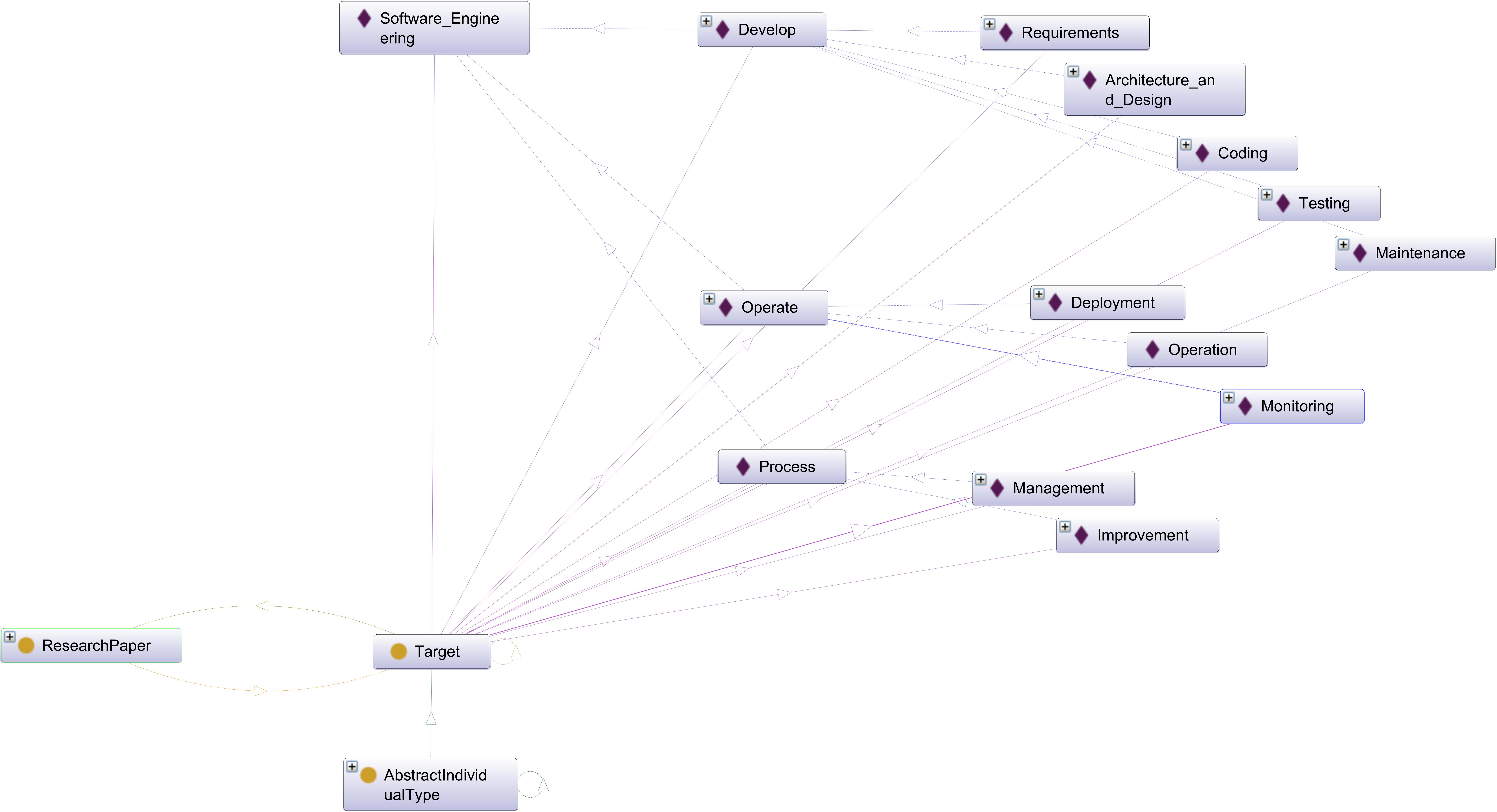} 
     \caption{Target dimension of the \textit{ai4se} ontology}
     \label{figai4setarget}
   \end{subfigure}% <--- This % sign is important, it prevents a horizontal space

   \caption{The \textit{ai4se} taxononmy defined as an ontology}\label{figai4seowl}
\end{figure}

To define this taxonomy more formally, the Web Ontology Language (OWL)~\citep{mcguinnessowl2004} was used, due to its association with the Semantic Web and the support offered by the Protégé tool~\citep{musenprotege2015} (see Figure~\ref{figai4seowl}). 
The \textit{ai4se} ontology definition is based on the gUFO ontology~\citep{almeidagufo2019} as a grounding and uses, in addition, the Computer Science Ontology (CSO~\citep{salatinocomputer2020}) to associate a general research topic with a research paper. 

A research paper itself is a mixin endurant and its research contribution is classified using abstract individual types (see Figure~\ref{figai4seontology}). The target dimension of the \textit{ai4se} taxonomy as described above is illustrated in Figure~\ref{figai4setarget}. Depending on the granularity of research paper's research target, it can address individual targets, such as \textit{requirements}, as well as larger activities, such as software \textit{development} or even \textit{software engineering} as a whole. 
The \textit{samplePaper} classified as an illustration in Figure~\ref{figai4sesample} has the \textit{CSO} research topic \textit{DevOps}. It focuses on \textit{understanding} and \textit{improving} software \textit{deployment}. It uses \textit{deep learning} and \textit{agentic AI}, and provides an \textit{AI-driven partial automation}. 

However, as the ontology lends itself to interactive elaboration rather than graphical representation, this paper uses a different tabular representation to depict the classification of the research papers, as shown in Figure~\ref{figAI4SE}. Interested readers can, however, explore the classification of research papers directly in the ontology~\citep{schieferdecker_ai4se_2025}, to which they can also contribute. The \textit{ai4se} ontology can be used not only to understand the concepts in the research field of AI for SE better, but also to explore research related to a specific aspect, such as all papers on agentic AI for SE, using SPARQL queries as shown in Listing~1 and~2\footnote{SPARQL is the standard query language and protocol for linked open data on the Web as defined by RDF triples. The OWL ontology \textit{ai4se} is already encoded in RDF triples, making it immediately queryable with SPARQL.}. 
\textit{ai4se} has been defined as an ontology because,  although taxonomies have been proposed for every area of SE knowledge within the SWEBOK~\citep{swebok2024} as discussed in~\citet{usman_taxonomies_2017}, neither SWEBOK nor any other source defines a comprehensive SE ontology that could be used to classify research results for the purpose of this paper. For example, while the CSO is quite detailed, it does not cover the knowledge areas of SWEBOK. For example, it lacks the concept of 'runtime monitoring' of software components and includes 'object-oriented design' at the top level, even though this is just one type of software design.

\begin{lstlisting}[language=SPARQL, caption={All papers on AI-assisted options automation.}, label={listingAIassisted}]
PREFIX rdf: <http://www.w3.org/1999/02/22-rdf-syntax-ns#>
PREFIX ai4se: <http://purl.org/ai4se/ontology#>
SELECT ?paper
WHERE {
 ?paper ai4se:hasLevel ai4se:AI-assisted_options .
}
\end{lstlisting}

\begin{lstlisting}[language=SPARQL, caption={All software engineering targets addressed by research papers.}, label={listingTargets}]
PREFIX rdf: <http://www.w3.org/1999/02/22-rdf-syntax-ns#>
PREFIX ai4se: <http://purl.org/ai4se/ontology#>
SELECT DISTINCT ?target
WHERE {
 ?paper rdf:type ai4se:ResearchPaper .
 ?paper ai4se:hasTarget ?target .
}
\end{lstlisting}

	\begin{table}[!ht]
			\centering 
			\footnotesize
			\rowcolors{1}{AntiqueWhite!30}{WhiteSmoke!30}
			\arrayrulecolor{Wheat}
			\begin{tabular}{|@{\hspace{2pt}}>{\raggedright\arraybackslash}p{0.1\linewidth}|@{\hspace{2pt}}>{\raggedright\arraybackslash}p{0.6\linewidth}|@{\hspace{2pt}}>{\raggedright\arraybackslash}p{0.03\linewidth}|@{\hspace{2pt}}>{\raggedright\arraybackslash}p{0.037\linewidth}|@{\hspace{2pt}}>{\raggedright\arraybackslash}p{0.037\linewidth}|@{\hspace{2pt}}>{\raggedright\arraybackslash}p{0.03\linewidth}|}
				\hline
				\textbf{\rotatebox{90}{Paper}} & \textbf{\rotatebox{90}{Summary}} & \textbf{\rotatebox{90}{Purpose}} & \textbf{\rotatebox{90}{Target}} & \textbf{\rotatebox{90}{Type}} & \textbf{\rotatebox{90}{Level}}\\
				\hline
				
				\citet{lin_traceability_2021}& T-BERT generates trace links between source code and natural language artefacts. It represents a significant advancement in automated software traceability, offering improved accuracy, efficiency, and practicality. & P2 & S1 & T2.c & L3 \\
				\hline
				
				\citet{korner_transferring_2014} & Natural language processing (NLP) approach for improving requirements engineering in the automotive industry: It identifies flaws, such as ambiguous words, incomplete process descriptions, or incorrect quantifiers.& P3 & S1.a & T1 & L2 \\
				\hline
				
				 \citet{perini_machine_2012}& CBRank method for prioritising software requirements: It uses stakeholder input and requirement attributes to generate an optimal ranking and enables adaptive elicitation by refining the prioritisation based on real-time feedback.& P3 & S1.a, S2.a, S3.a & T1, T2.b & L2 \\
				\hline
				
				 \citet{bhat_adex_2019}& ADeX is a tool designed to automate the curation of architectural design decisions in software development. It helps automate the management of architectural decisions, reducing documentation effort while improving decision accuracy. & P1, P3 & S1.b, S2.a & T1, T2.b & L3 \\
				\hline
				
				 \citet{bird_taking_2023} & Explores the early experiences of developers using the AI-augmented coding assistant GitHub Copilot. It may expand to debugging, code review, and software maintenance, requiring better trust and provenance tracking. & P1, P2 & S1.c & T3 & L3\\
				\hline

				 \citet{svyatkovskiy_intellicode_2020}& IntelliCode Compose is a code completion tool that uses a generative transformer model to predict sequences of code tokens, including entire lines of code. It has been trained on a massive dataset of 1.2 billion lines of code across multiple programming languages. & P2, P3 & S1.c & T2.a, T2.c, T3 & L3\\
				\hline
				
				 \citet{bader_getafix_2019}& Getafix is an automated bug-fixing tool that learns from past, human-written fixes to suggest human-like fixes for static analysis warnings. & P2, P3 & S1.c, S1.d & T2.a, T2.b & L2 \\
				\hline
				
				\citet{guo_graphcodebert_2020}& GraphCodeBERT improves code understanding by incorporating dataflow graphs into the BERT transformer architecture. Data flow graphs represent dependencies between code variables and improve semantic understanding over treating code as a simple sequence of tokens. & P1 & S1.c, S1.d, S1.e & T2 & L1 \\
				\hline				
		\end{tabular}
		\caption{AI-assisted software engineering I (sorted by application targets)} \label{figCurrentAI4SEa}
		\end{table}

	\begin{table}[!ht]
	\centering 
	\footnotesize
	\rowcolors{1}{AntiqueWhite!30}{WhiteSmoke!30}
	\arrayrulecolor{Wheat}
	\begin{tabular}{|@{\hspace{2pt}}>{\raggedright\arraybackslash}p{0.1\linewidth}|@{\hspace{2pt}}>{\raggedright\arraybackslash}p{0.6\linewidth}|@{\hspace{2pt}}>{\raggedright\arraybackslash}p{0.03\linewidth}|@{\hspace{2pt}}>{\raggedright\arraybackslash}p{0.037\linewidth}|@{\hspace{2pt}}>{\raggedright\arraybackslash}p{0.037\linewidth}|@{\hspace{2pt}}>{\raggedright\arraybackslash}p{0.03\linewidth}|}
		\hline
		\textbf{\rotatebox{90}{Paper}} & \textbf{\rotatebox{90}{Summary}} & \textbf{\rotatebox{90}{Purpose}} & \textbf{\rotatebox{90}{Target}} & \textbf{\rotatebox{90}{Type}} & \textbf{\rotatebox{90}{Level}}\\
		\hline

				 \citet{gupta_intelligent_2018}& DeepCodeReviewer is an automatic code analysis system that learns from historical peer reviews to identify and suggest relevant reviews for code snippets. & P1, P3 & S1.c, S1.d & T2 & L2 \\
				\hline
				
				 \citet{drain_generating_2021} & Introduces DeepDebug or automated bug detection and fixing in Java: It generates non-deletion fixes that modify instead of just deleting code & P1, P2 & S1.c, S1.e & T3 & L3\\
				\hline

				 \citet{tufano_generating_2022}& Approach improves code review automation with DL by working with pre-trained models on raw source code and English text.& P1 & S1.c, S1.d & T3 & L2 \\
				\hline
				
				 \citet{tatineni_integrating_2024}& The book discusses the application of AI beside others for infrastructure monitoring and anomaly detection, predictive maintenance, and build pipeline optimization. & P1, P3 & S1.c, S1.d, S2.a, S3.b & T2.a, T2.b & L3 \\
				\hline

				 \citet{bouzenia_repairagent_2024}& Presents RepairAgent for automated program repair that allows to autonomously plan and execute actions, such as gathering information, searching for code snippets, and validating fixes, by interacting with a set of specifically designed tools. & P3 & S1.c, S2.a & T4 & L3 \\
				\hline
				
				 \citet{lenz_linking_2013}& Approach links different types of testing information to help automate the generation of equivalence classes for testing. The linking supports tasks such as test reduction, prioritisation and selection.& P1, P2, P3 & S1.d & T2.a, T2.b & L2 \\
				\hline
				
				 \citet{zhao_clustering_bayesian_2015}& Presents a hybrid approach to test case prioritization for regression testing aiming to minimise testing costs. Aims to enhance software testing efficiency by optimising test case execution order.& P3 & S1.d, S1.e & T1, T2.a, T2.b & L2 \\
				\hline
				
				 \citet{spieker_reinforcement_2017}& Introduces Retecs for automatic test case prioritisation and selection in continuous integration. It uses test case meta-data to adapt to changes in the environment and prioritise error-prone test cases. & P3 & S1.d, S2.a & T2.b, T2.c & L3 \\
				\hline
				
				 \citet{ceran_prediction_2023}& Explores machine learning-based ensemble methods to predict software quality more accurately than previous approaches & P1 & S1 & T2.b & L3\\
				\hline
				
			\end{tabular}
	\caption{AI-assisted software engineering II (sorted by application targets)} \label{figCurrentAI4SEb}
	\end{table}

\subsection{Literature Review and Classification with the \textit{ai4se} Taxonomy}

The fidelity of the new taxonomy \textit{ai4se} (Figure~\ref{figAI4SE}) is verified by reviewing and classifying highly cited publications, but also recent publications to demonstrate the applicability of this taxonomy.  A lightweight systematic literature review (SLR), as described in~\citet{stapicperforming2012}, was conducted to analyse related research. This SLR protocol was followed:

\begin{itemize}
	\item \textbf{Review title}: SLR on evaluating the status of AI for SE research by applying the \textit{ai4se} taxonomy.
	\item \textbf{Objectives of the review}: \vspace{-0.2cm}
		\begin{enumerate}
			\item To determine research an AI for SE
			\item To test the validity of the \textit{ai4se} taxonomy with a research selection.
			\item To determine a classification of the selected research.
		\end{enumerate}\vspace{-0.2cm}
	\item \textbf{Research questions}:\vspace{-0.2cm}
		\begin{itemize}
			\item RQ1: What is the current status of research on AI for SE and for MDSE?
			\item RQ2: Are the dimensions of the \textit{ai4se} taxonomy suitable for classifying current AI for SE research results?
			\item RQ3: Are all dimensions and facets of the taxonomy represented in research work?
			\item RQ4: What are typical purposes addressed, targets tackled, types used, and levels of automation with AI reached?
		\end{itemize}\vspace{-0.2cm}
	\item \textbf{Database}: An automated search on dblp, ACM DL, and IEEE Xplore of SE and AI related research papers published between 2020 and 2025. Snowballing extension of the search results with Research Rabbit, Citation Gecko, and Google Scholar. Consolidation of the search results with Zotero. 
	\item \textbf{Inclusion criteria}: \vspace{-0.2cm}
		\begin{itemize}
			\item Original research papers in English.
			\item Online available.
			\item Research on AI for SE.
		\end{itemize}\vspace{-0.2cm}
	\item \textbf{Exclusion criteria}: \vspace{-0.2cm}
		\begin{itemize}
			\item Research on SE for AI.
		\end{itemize}\vspace{-0.2cm}
	\item \textbf{Selection process}:\vspace{-0.2cm}
		\begin{enumerate}
			\item Title and abstract screening for the pre-selection of primary research reporting original research contributions by use of the concept map resulting from the \textit{ai4se} ontology including synonyms.
			\item Full text review and assessment of the research contributions for the final selection of primary research.
			\item Tools: Python for text analysis and post-processing of finally selected research, supported by MS Visual Studio, Google AI Studio, and LibreOffice.
		\end{enumerate}\vspace{-0.2cm}
	\item \textbf{Synthesis process}: \vspace{-0.2cm}
	\begin{itemize}
		\item Review of the finally selected research and inclusion in the \textit{ai4se} ontology.
	\end{itemize}\vspace{-0.2cm}
\end{itemize}

A total of 229~studies on AI for SE were identified. Of these, 32~were primarily SLRs, 46~papers described (elements of) research roadmaps for (aspects of) AI for SE, and 10 papers presented taxonomies for (aspects of) AI for SE.  Twelve SLRs, eight roadmaps and seven taxonomies are summarised and compared in Tables~\ref{figSLRAI4SE}, \ref{figRoadmapsAI4SE} and~\ref{figTaxonomyAI4SE} respectively, due to their relation to the \textit{ai4se} taxonomy. A further 82~papers present original methods, techniques, and tools for AI for SE excluding AI for MDSE. Of these, 17~were finally selected and classified. For research on AI for MDSE, see Section~\ref{BigModels} for details of the 15~research papers selected from a total of 59~research items on AI for MDSE. Please note that the SLR results are available online, and that the \textit{ai4se} taxonomy will be extended in the future to include further original research papers.

The selected original research on AI for SE in general is presented in Table~\ref{figCurrentAI4SEa}, \ref{figCurrentAI4SEa}. Additional details are given here: 
For \textbf{requirements},~\citet{korner_transferring_2014} presents an AI-based automation approach to requirements engineering that begins by converting natural language into an Eclipse Modelling Framework (EMF) model. It then applies linguistic rules to identify errors, such as ambiguities or incorrect quantifiers, and provides suggestions for requirements analysts to make final decisions. This approach supports the entire requirements elicitation and change process. Another \textbf{requirements} \textit{improvement} is given in~\citet{perini_machine_2012} presenting the prioritisation of requirements by combining the preferences of project stakeholders with approximations of the order of requirements computed by ML techniques.

For \textbf{architecture and design},~\citet{bhat_adex_2019} describes the automated curation of design decisions to support architectural decision-making. It helps software architects by organizing and recommending design decisions based on previous cases and contextual information of the current project. By leveraging existing design knowledge, the approach analyses historical data and design choices to improve the quality and consistency of architectural designs.

The list of publications on AI-assisted \textbf{coding}, \textbf{testing}, and \textbf{maintenance} is huge. Major developments are for example described in~\citet{gupta_intelligent_2018} for \textbf{code} \textit{understanding}. It introduces DeepCodeReviewer that uses deep learning to recommend code reviews for common issues based on historical peer reviews. It assesses the relevance of reviews to specific code snippets, suggests appropriate reviews from a repository of common feedback, and improves code reviews by focusing on defect detection. \citet{guo_graphcodebert_2020} describes GraphCodeBERT, a pre-trained model for programming languages that incorporates data flow semantics rather than just code syntax. GraphCodeBERT demonstrates its performance both in code understanding for code search and clone detection, and in code generation and improvement through code translation and code refinement. \citet{ceran_prediction_2023} presents a study focused on predicting software quality using defect density as a key feature representing quality to achieve higher accuracy in software quality prediction compared to previous studies. The research shows that data pre-processing, feature extraction and the application of ML algorithms significantly improve prediction accuracy. \citet{bird_taking_2023} discusses early experiences of developers using GitHub Copilot, which uses a language model trained on source code. Guided by Copilot, developers can write code faster than a human colleague, potentially accelerating development. Three empirical studies with Copilot highlight the different ways developers use Copilot, the challenges they face, the evolving role of code review, and the potential impact of pair programming with AI on software development. \citet{svyatkovskiy_intellicode_2020} discusses IntelliCode Compose, a multilingual code completion tool that predicts entire sequences of code tokens up to full lines of code. The generative transformer model has been trained on 1.2 billion lines of Python, C\#, JavaScript and TypeScript code. \citet{bader_getafix_2019} presents Getafix, a tool for fixing common bugs by learning from previous human-written fixes. It uses hierarchical clustering to group bug fix patterns into a hierarchy from general to specific, and a ranking system based on the context of the code change to suggest the most appropriate fix. Another debugging approach, DeepDebug, is presented in~\citet{drain_generating_2021}, which has been trained by mining GitHub repositories to detect and fix bugs in Java methods. 

Since different software \textbf{testing} techniques are complementary, reveal different types of defects and test different aspects of a program,~\citet{lenz_linking_2013} presents an ML-based approach to link test results from different techniques, to cluster test data based on functional similarities, and to generate classifiers according to test objectives, which can be used for test case selection and prioritisation. \citet{tufano_generating_2022} discusses a test generation approach for writing unit test cases by generating assert statements. The approach uses a transformer model that was first pre-trained on an English text corpus, further semi-supervisedly trained on a large source code corpus, and finally fine-tuned for the task of generating assert statements for unit tests. The assert statements are accurate and increase test coverage. \citet{zhao_clustering_bayesian_2015} discusses test case prioritisation based on source code changes, software quality metrics, test coverage data, and code coverage-based clustering. It reduces the impact of similar test cases covering the same code and improves fault detection performance.
			
For \textbf{deployments},~\citet{tatineni_integrating_2024} explores the role of ML in DevOps for intelligent release management. It suggests besides other things combining continuous monitoring, predictions of the likelihood of deployment failures, root cause analysis, and pipeline optimisation to reduce deployment failures and improve release management efficiency and software quality. 

For the SE \textbf{process},~\citet{spieker_reinforcement_2017} introduces Retecs, a method for automatically learning test case selection and prioritisation in continuous integration, aimed at minimising the time between code commits and developer feedback on failed tests. Retecs uses reinforcement learning to select and prioritise test cases based on their execution time, previous execution history and failure rates. It effectively learns to prioritise error-prone test cases by following a reward function and analysing past CI cycles. \citet{lin_traceability_2021} presents the T-BERT framework for generating trace links between source code and natural language artefacts such as requirements or code issues. It demonstrates superior accuracy and efficiency for software traceability, especially in data-limited environments.

Overall, the majority of the presented selection of recent research focuses on improving the software engineering process, either by improving coding, requirements engineering, testing, or the overall processes. A smaller number of approaches aim to improve the understanding of software engineering artefacts. There are also methods that focus on generating code or code-related content. A mixture of AI types are used, including symbolic, statistical, ML, DL and GenAI, as well as hybrid approaches. Many of the AI approaches offer AI-assisted selection, i.e., they provide suggestions, recommendations and rankings that can help developers, but require a developer to make the final decision on what to do. AI-based partial automation is also a common level, where tools automate part of the process, but may require a human to review or further integrate the results. On the whole, the dimensions of the \textit{ai4se} taxonomy are adequately covered, although not every possible combination has yet been addressed — it may even be impossible to do so. For example, the prospect of full automation driven by AI using only symbolic AI is unlikely, since the nuances of SE can only be effectively managed with more advanced non-symbolic AI. However, as AI for SE is a very active area of research, further research results will emerge that make use of the open combinations of the taxonomy dimensions. To accommodate these results, the taxonomy is available online~\citep{schieferdecker_ai4se_2025} and will be continuously updated.

With regard to the research questions provided for the SLR, which formed the basis for the development of the \textit{ai4se} taxonomy, the answers are as follows: The SLR revealed current research on AI for SE and for MDSE as shown in Table~\ref{figCurrentAI4SEa},~\ref{figCurrentAI4SEb} and Table~\ref{figCurrentAI4MDSEa}, ~\ref{figCurrentAI4MDSEb}. Software development and model-driven software design are the primary focus of AI support in SE, but also other SE and MDSE activities are being studied. The \textit{ai4se} taxonomy can classify the identified primary research, that address all dimensions and their facets, but not general purpose AI (T5) or AI-driven full automation (L4). Subsymbolic AI (T2) and generative AI (T3) are often used, agentic AI (T4) is on a rise. The automation levels reached are most often AI-assisted option (L1) and AI-assisted selection (L2).

%%%%%%%%%%%%%%%%
\subsection{Perspective on AI-Assisted Model-Driven Software Engineering}\label{BigModels}

The substantial increase in the utilisation of AI for SE, as described in the previous section, is in parallel with a notable rise in the application of AI to MDSE in the following areas: model creation and elicitation; model completion, refinement and repair; intelligent model transformations; model validation and verification; reverse engineering and model discovery; and model-based testing. Using AI in MDSE enables the management of unprecedented complexity, improving the quality and consistency of software projects. These developments can also be explained by the novel 'model naturalness hypothesis', which extends the naturalness hypothesis~\citep{allamanis_survey_2017} to modelling artefacts in software engineering: \\

\begin{flushright}
\begin{minipage}[t]{0.9\textwidth} 
\begin{spacing}{1} 
\textbf{The model naturalness hypothesis}\\

Software is a form of human communication as well as a form of human-computer communication; \textbf{model corpora}, software corpora, and natural language corpora have similar statistical properties; the properties of model corpora can be employed to develop more efficacious software engineering tools.
\end{spacing}
\end{minipage}
\end{flushright}

Readers familiar with software modelling techniques will not be surprised by this hypothesis. Like software code, software models are expressed in (semi-)formally defined languages with precise, often graphical, syntax and semantics. In this respect, software models are similar to software code. However, they are often only used in the initial phases of software design to sketch the desired solution. Consequently, they are used to visualise code and are not stored in a machine-readable format for reuse and further processing. In contrast, however, software models can facilitate human-human and human-computer communication if they are treated as machine-processable, high-level software artefacts rather than images (see also~\citet{babur_models_2018}). In this case, models can form corpora that can be analysed automatically and used to train AI. The model naturalness hypothesis has therefore been made explicit. 
Indeed, the advent of big code and AI has opened up new avenues for integrating AI into model-driven approaches. As demonstrated by~\citet{hamilton2017representation}, the intrinsic formal graph structures of SE models can also be leveraged to facilitate the preparation of models for gaining AI advantages in MDSE. This is analogous to the considerations in~\citet{pudari_copilot_2023} on the necessity to elevate abstraction levels in language models to enhance the efficacy of AI-assisted automated coding. 

	\begin{table}[!ht]
			\centering 
			\footnotesize
			\rowcolors{1}{AntiqueWhite!30}{WhiteSmoke!30}
			\arrayrulecolor{Wheat}
			\begin{tabular}{|@{\hspace{2pt}}>{\raggedright\arraybackslash}p{0.1\linewidth}|@{\hspace{2pt}}>{\raggedright\arraybackslash}p{0.6\linewidth}|@{\hspace{2pt}}>{\raggedright\arraybackslash}p{0.03\linewidth}|@{\hspace{2pt}}>{\raggedright\arraybackslash}p{0.037\linewidth}|@{\hspace{2pt}}>{\raggedright\arraybackslash}p{0.037\linewidth}|@{\hspace{2pt}}>{\raggedright\arraybackslash}p{0.03\linewidth}|}
				\hline
				\textbf{\rotatebox{90}{Paper}} & \textbf{\rotatebox{90}{Summary}} & \textbf{\rotatebox{90}{Purpose}} & \textbf{\rotatebox{90}{Target}} & \textbf{\rotatebox{90}{Type}} & \textbf{\rotatebox{90}{Level}}\\
				\hline			
				
				 \citet{babur_models_2018}& This paper presents quantitative evidence from academia and industry regarding the increased use of models. It suggests using models as data and applying machine learning techniques to improve our understanding of them. & P1 & S1.m& T2.b & L0\\
				\hline
				
				 \citet{kulkarni_toward_2023}& Discusses NLP-driven model generation for more efficient model construction. Presents "purposive" meta-models that guide the interactions between domain experts and GenAI to generate effective prompts for model construction. & P2 & S1.a.m, S1.b.m & T1, T3 & L3\\
				\hline
				
				 \citet{chen_design_2024}& Discusses the NLP-driven domain modelling by evaluating different LLMs and various prompt engineering techniques. & P2 & S1.a.m, S1.b.m & T3 & L3\\
				\hline
				
				 \citet{adhikari_simima_2024}& This paper introduces SimIMA, an intelligent modelling assistant for Simulink. It uses association rule mining and model clone detection to support the modeller in developing models. & P3 & S1.b.m& T2.b & L1\\
				\hline
				
				 \citet{saini_automated_2022}& This paper proposes an approach to bot-modeller interaction. It combines an incremental learning strategy with the discovery of alternative configurations, improving the accuracy of the bot's suggestions through the analysis of domain modelling decisions over time.& P2, P3 & S1.b.m& T2.b & L1\\
				\hline
						
				 \citet{khalilipour_machine_2022}& This paper classifies Ecore metamodels by using their structural and textual information to improve automated model classification for model management. & P1 & S1.b.m& T2.a & L0\\
				\hline
				
				 \citet{hartmann_next_2017}& This study examines an approach to improving domain modelling by incorporating micro-learning into the modelling process.& P3 & S1.b.m& T2.a & L0\\
				\hline

	\end{tabular}
	\caption{AI-augmented model-driven software engineering I (sorted by application targets)} \label{figCurrentAI4MDSEa}
	\end{table}
	\begin{table}[!ht]
	\centering 
	\footnotesize
	\rowcolors{1}{AntiqueWhite!30}{WhiteSmoke!30}
	\arrayrulecolor{Wheat}
	\begin{tabular}{|@{\hspace{2pt}}>{\raggedright\arraybackslash}p{0.1\linewidth}|@{\hspace{2pt}}>{\raggedright\arraybackslash}p{0.6\linewidth}|@{\hspace{2pt}}>{\raggedright\arraybackslash}p{0.03\linewidth}|@{\hspace{2pt}}>{\raggedright\arraybackslash}p{0.037\linewidth}|@{\hspace{2pt}}>{\raggedright\arraybackslash}p{0.037\linewidth}|@{\hspace{2pt}}>{\raggedright\arraybackslash}p{0.03\linewidth}|}
		\hline
		\textbf{\rotatebox{90}{Paper}} & \textbf{\rotatebox{90}{Summary}} & \textbf{\rotatebox{90}{Purpose}} & \textbf{\rotatebox{90}{Target}} & \textbf{\rotatebox{90}{Type}} & \textbf{\rotatebox{90}{Level}}\\
		\hline

				 \citet{baki_multi-step_2016} & Presents an approach to automatically generate model transformations from examples. The approach is demonstrated through experimental evaluation on seven diverse transformation problems. & P1, P2 & S1.b.m & T1, T2.a, T2.b & L3\\
				\hline
				
				 \citet{mangaroliya_classification_2020}& Explores the automated classification of UML class diagrams into forward-engineered and reverse-engineered diagrams. It provides a dataset with ground truth for the classification of class diagrams. & P1, P3 & S1.b.m & T2.b & L2\\
				\hline
								
				 \citet{acher_generative_2023}& Presents an experience report detailing the use of LLMs to analyse and transform software variants represented in various formalisms (Java, UML, Graph Markup Language (GraphML), statecharts).& P1, P2 & S1.b.m, S1.c.m & T3, T4 & L2\\
				\hline

				 \citet{eramo_architecture_2024}& Presents a software architecture for the AI-augmented automation of DevOps pipelines to address both architectural and technical challenges associated with the development and operation of complex systems.& P1 & S1.b.m, S1.e.m & T2.c & L2\\
				\hline
				
				 \citet{babur_statistical_2016}& Proposes an approach for statistically analysing large collections of models within the field of MDSE. The approach demonstrates its effectiveness in domain analysis and repository management, and shows the potential for applications such as model clone detection.& P1 & S1.b.m, S1.e.m& T2.a, T2.b & L1 \\
				\hline
				
				 \citet{xue_automating_2023}& This work proposes Code Generation By Example (CGBE), a code generation approach that has been learned from examples for UML to Java code generation. & P2 & S1.c.m& T1 & L1\\
				\hline
				
				 \citet{lin_soen-101_2024}& Introduces FlowGen, a code generation framework that leverages multiple LLM agents to emulate different software process models (Waterfall, Test-Driven development (TDD), and Scrum). On standard benchmarks, the Scrum-based FlowGenScrum consistently outperforms other models and baselines. & P2 & S1.c.m, S1.d.m, S1.e.m, S3.1.m, S3.2.m & T3, T4 & L3\\
				\hline
				
				 \citet{tamizharasi_novel_2024}& This study presents an approach to deriving feature models in UML from historical source code, which can then be used as a basis for generating and prioritising test cases.  & P2 & S1.d.m& T2.b & L2\\
				\hline

			\end{tabular}
	\caption{AI-augmented model-driven software engineering II (sorted by application targets)} \label{figCurrentAI4MDSEb}
	\end{table}
	
\normalsize

Foundational work on ground truths for developing and benchmarking AI applications in MDSE is discussed in~\citet{torcal_creating_2024,shcherban_multiclass_2021} and~\citet{mangaroliya_classification_2020}. \citet{torcal_creating_2024} presents the creation and validation of a new, high-quality ground truth dataset of UML diagrams. The curated dataset of 2626 unique and accurately labelled images across six UML diagram types goes beyond the Lindholmen dataset of modelling artefacts~\citep{robles_extensive_2017, robles_reflection_2023}. The new ground truth dataset was also presented for an improved approach to UML diagram classification. Also~\citet{shcherban_multiclass_2021} presents a new dataset of images, manually labelled into four UML diagram types (class, activity, sequence, use case) and a non-UML category, which is used to develop new DL approaches with higher precision to automate UML diagram classification. Another ground truth dataset is described in~\citet{mangaroliya_classification_2020}. 

The selected original research on AI for MDSE is categorized in Table~\ref{figCurrentAI4MDSEa} and~\ref{figCurrentAI4MDSEb}. Further details can be found here:
The majority of the original work on AI for MDSE is focuses on supporting the design of software: \citet{babur_models_2018} was one of the first studies to suggest treating models as data for training LLMs, enabling the application of ML techniques to MDSE. 

Another significant area of research concerns the use of AI to improve domain modelling. \citet{kulkarni_toward_2023} presents a methodology for AI-augmented MDSE as a continuous cycle of AI-generated model proposals by LLMs, which are then reviewed and refined by humans. This approach aims to bridge the gap between the natural language used by domain experts and the formal languages employed in MDSE. 
\citet{chen_automated_2023} investigates the feasibility of using LLMs to automate domain modelling in SE. It applies various prompt engineering techniques to a dataset of diverse domain modelling examples and finds that while LLMs show promising domain understanding and achieve relatively high precision in generating classes and attributes, their recall remains low, particularly for relationships. It concludes that full automation is currently impractical. 

\citet{saini_automated_2022} addresses the question of how to improve the modeller–bot interaction. The bot presented in the study suggests alternative model configurations based on a novel algorithm and updates the domain model based on modeller input. This addresses the limitations of existing automated methods. Evaluations demonstrate high accuracy and rapid performance. 
\citet{hartmann_next_2017} also examines the interaction between the modeller and the AI tool and propose a multistrategy learning approach for the gradual, goal-driven refinement of incomplete domain models provided by human experts. The system learns general rules from specific examples and refines or extends concepts in order to handle exceptions caused by model incompleteness.
An intelligent modelling assistant for Simulink is presented in~\citet{adhikari_simima_2024}. It offers two forms of data-driven guidance: SimGESTION provides real-time edit suggestions using ensemble learning, while SimXAMPLE recommends related models through clone detection.

Model management is central to MDSE and is therefore an active area of research. The methodology in~\citet{babur_statistical_2016} uses techniques from information retrieval, NLP, and ML to enable the efficient comparison, clustering and clone detection of model artefacts. 
The automated classification of UML class diagrams into forward-engineered and reverse-engineered diagrams presented in~\citet{mangaroliya_classification_2020} provides another perspective on model management that is crucial for understanding the evolution of software projects. 
Different to that,~\citet{khalilipour_machine_2022} focuses on Ecore metamodels and presents results showing that incorporating structural information significantly improves the accuracy with which models are classified, which in turn helps to improve the comparison, clustering, and clone detection of model artefacts. 

Model transformation has also been studied:~\citet{baki_multi-step_2016} presents a novel method for automatically generating model transformations from examples. It employs a three-step process: (1) analysing example model pairs to identify groups of similar transformations, (2) using genetic programming to learn basic rules for each group, and (3) refining these rules using simulated annealing to handle complex dependencies and value derivations. The method presented improves the efficiency and accuracy of learning complex model transformations.

Code generation supported by models and AI has also been a key research field: 
\citet{xue_automating_2023} aims to improve the synthesis of reusable and adaptable code generators in MDSE by reducing the manual effort typically required. The proposed approach, Code Generation By Example (CGBE), uses symbolic machine learning and tree-to-tree mappings to learn code generation from UML to Java.
\citet{lin_soen-101_2024} uses agentic AI where each agent embodies a specific development role such as requirement engineer, developer, or tester and collaborates via chain-of-thought prompting. The set of LLM agents emulate different software process models and are evaluated against four benchmarks. The impact of individual development activities on code quality is also investigated, revealing the critical role of testing and the beneficial effects of design and code review

Similarly, testing is an important area for applying improved generation methods from models supported by AI.
The test case generation and prioritisation presented in~\citet{tamizharasi_novel_2024} addresses the challenges of resource constraints, high costs and inefficiencies in traditional test development methods.

Case studies on AI-augmented MDSE have also been published. 
\citet{acher_generative_2023} explores the use of LLMs to aid in the complex process of re-engineering existing software variants into a software product line (SPL) by identifying features to represent variability and commonalities. It describes the methodology used to prompt the LLMs, the results obtained, and the limitations encountered. There is a potential in assisting domain analysts and developers in transforming software variants into SPLs, but LLMs are limited in managing large variants or complex structures.
\citet{eramo_architecture_2024} presents a novel software architecture designed to address the increasing complexity of modern system development within a DevOps framework. The architecture integrates MDSE and AI techniques to enhance the entire software lifecycle, from requirements engineering to monitoring and maintenance. The architecture is evaluated through multiple industrial case studies.

All papers share the common view that the strengths of AI and MDSE should be combined to further improve the quality of software products and/or processes. While acknowledging the potential of AI-supported MDSE, the papers also highlight the limitations of LLMs, their sensitivity to prompt variations, their limited ability to handle complex inputs, and the risk of inaccuracies. There is a consensus that a fully automated approach is not yet feasible and that human review remains necessary to counteract inaccuracies introduced by AI. UML is seen in these papers as a common modelling language, both for creating models and as an input to AI models. There is also an expectation that AI techniques will support better model quality and improve software reuse by identifying and adapting components from previous projects. This will be supported by the use of model matching and clone detection techniques.

In general, the potential of using AI for MDSE can be summarised as follows: MDSE provides the formal structure and 'guardrails' that are required for AI to operate effectively and reliably. In turn, AI provides the intelligence and flexibility needed to automate the most challenging and creative stages of the MDSE lifecycle. To realise these prospects in industrial SE practices, the following opportunities require broader uptake:
\begin{enumerate}
	\item The provision of large amounts of \textbf{top-down models} on coding platforms.
	\item The generation of \textbf{bottom-up models} from big code.
	\item The formation of \textbf{Big Models} as a basis for more advanced empirical MDSE and further improvements of MDSE. 
	\item The \textbf{application of AI methods} on big models and for big models.
	\item The shift towards the paradigm of \textbf{pair modelling} in industrial SE practice will altogether turn SE into the \textbf{next generation of software engineering}.
\end{enumerate}

Regarding \textbf{opportunity (1)},~\citet{storrle_index_2014} presented the first entities of the SEMI Software Engineering Models Index, a catalogue of model repositories, and invited further contributions to SEMI. \citet{hebig_quest_2016} used another approach of mining GitHub for projects including UML models, which could well be combined with the SEMI initiative. As a result of numerous efforts towards SE model collections, the Lindholmen dataset was developed~\citet{robles_extensive_2017} and reviewed in~\citet{robles_reflection_2023}. For recent work on ground truth datasets see Table~\ref{figCurrentAI4MDSEa} and~\ref{figCurrentAI4MDSEa}.

In view of \textbf{opportunity (2)}, extracting representational models from code is in fact the opposite of using models to generate code more efficiently. Since the design, specification, and maintenance of such models can be complex and time-consuming, various techniques have been developed to extract models from code and/or execution traces. These techniques include the extraction of all types of models including for example the extraction of information models, see e.g.~\citet{burson_program_1990,murphy1996lightweight}, of structural models, see e.g.~\citet{kazman1998requirements,guo1999software}, and of behavioural models, see e.g.~\citet{corbett_bandera_2000,lo_automatic_2009}. The process of extracting models from code and/or execution traces offers the advantage of retrieving models that are up-to-date with the code/traces, but it also carries the potential disadvantage of mismatch with the model representation and abstraction requirements. 

With regard to \textbf{opportunity (3)}, there are initial studies on big models such as~\citet{ho-quang_practices_2017} which explores the increasing role of modelling, especially in safety-critical software development. It surveyed a range of projects utilising the UML and identified collaboration as the primary rationale for employing models. This is because models facilitate communication and planning for joint implementation efforts within teams, including those who are not directly involved in modelling or are new to the team.

For \textbf{opportunity (4)}, Table~\ref{figCurrentAI4MDSEa}, ~\ref{figCurrentAI4MDSEb} elaborates the current state of research in more detail.

Finally, regarding  \textbf{opportunity (5)}, the pair modelling paradigm, initially proposed during the peak of top-down MDSE~\citep{kamthan2008pair}, is undergoing a significant evolution. In the era of AI-augmented software engineering (SE), it is emerging as a natural progression from AI-assisted pair programming~\citep{bipp2008pair, dakhel_github_2023}, which has already become a prominent application of AI in the field. 

Driven by the development of foundation models tailored for MDSE (analogous to LLMs trained on source code), AI-supported pair modelling has the potential to become a central methodology. In this setup, AI-powered tools and agents act as collaborative partners. Drawing from pair programming dynamics, this collaboration relies on two roles: the driver and the navigator (or observer). The driver actively constructs or refines software artefacts, including models. Meanwhile, the navigator continuously reviews the work, focusing on systemic and strategic concerns to identify improvements and preempt future issues.

This separation of concerns allows the driver to focus on immediate implementation without losing sight of overarching architectural goals. Crucially, the human engineer and the AI agent can dynamically switch between these roles. The exact interplay depends heavily on the AI tool's capabilities and the level of automation it can reliably provide to the software engineer.

\section{Related Work}\label{Related}

As this paper's main contribution is the \textit{ai4se} taxonomy, which was developed through a literature review of AI for SE (including MDSE), the related work section reviews literature survey approaches and taxonomy developments in the emerging field of AI applications in SE. Classifying research papers within this taxonomy also led to a discussion of opportunities in AI for MDSE. Therefore, the related work section also covers roadmaps on this topic.

Systematic literature reviews are used to explore the status of a research field by examining the results of published research presented in peer-reviewed papers that are available online. Numerous SLRs on AI for SE have been published. While some categorise the reviewed research, none have developed a taxonomy for classifying the field of AI for SE. Table~\ref{figSLRAI4SE} presents a selection of recent SLRs on AI for SE.  

	\begin{table}[!ht]
			\centering 
			\footnotesize
			\rowcolors{1}{AntiqueWhite!30}{WhiteSmoke!30}
			\arrayrulecolor{Wheat}
			\begin{tabular}{|@{\hspace{2pt}}>{\raggedright\arraybackslash}p{0.1\linewidth}|@{\hspace{2pt}}>{\raggedright\arraybackslash}p{0.45\linewidth}|@{\hspace{2pt}}>{\raggedright\arraybackslash}p{0.45\linewidth}|}
				\hline
				\textbf{Paper} & \textbf{Summary} & \textbf{Difference to \textit{ai4se}} SLR\\
				\hline
				
				\citet{alenezi_ai-driven_2025}& This paper focuses on the benefits of AI for SE, such as increased productivity, improved code quality and faster development, and on key impact areas such as automated coding, intelligent debugging and predictive maintenance. The paper supports its findings with case studies and theoretical models from various activities in SE. & SLR with focus on case studies.\\
				\hline
				
				\citet{durrani_impact_2025} & This study provides a quantitative assessment of the impact of AI in SE phases. It suggests that AI can improve the accuracy of planning and requirement engineering, as well as enhancing the efficiency of the latter and the software design.& Follow-up quantitative analysis of a SLR.\\
				\hline
				
				\citet{hou_large_2024} & This study analyses LLMs for SE by characterising their features, reviewing data curation and training processes, and evaluating their performance and success in application to SE.& SLR on LLMs for SE.\\
				\hline
				
				\citet{mohammad_challenges_2024} & This study examines the main challenges of applying AI to project management, particularly during the planning stage. & SLR on AI for project planning.\\
				\hline
				
				\citet{zhang_survey_2024} & This paper provides an overview of the general workflow of learning-based automated program repair techniques. It outlines related empirical studies and provides evaluation metrics.& SLR on AI for program repair.\\
				\hline
				
				\citet{durrani_decade_2024} & This paper aims at identifying the AI techniques most often used in software development and their effects on the accuracy and efficiency of software development activities. & SLR on software development. \\
				\hline
							
				\citet{wang_software_2024} & This paper investigates the software testing tasks for which LLMs are typically employed, the most prevalent LLMs, the types of prompt engineering employed alongside these LLMs, and the associated techniques. & SLR on LLMs for testing.\\
				\hline

				\citet{kumar_current_2023} & This study analysis the advances of AI for SE, the potentials for future
development as well as the risks of AI application to SE. & SLR combined with qualitative expert interviews.\\
				\hline
				
				\citet{wong_natural_2023} & This paper provides a review of transformer-based LLMs that have been trained using big code and are used for NLP techniques in AI-assisted programming. & SLR on NLP utilization for programming. \\
				\hline
				
				\citet{wang_machinedeep_2023} & This study examines the complexities involved in applying ML/DL solutions to SE.
It also explores how these issues affect the reproducibility and replicability of ML/DL applications in SE. & SLR on ML/DL for SE.\\
				\hline
				
				\citet{mohammadkhani_systematic_2023} & This study explores the extent to which explainable AI has been investigated within the SE community. Software maintenance, and defect prediction in particular, is the area of SE that has received the most attention in terms of the stages and tasks being studied.& SLR on explainable AI for SE.\\
				\hline
				
				\citet{fan_large_2023} & This paper reviews the use of LLMs in SE and highlights the important role of a hybrid approach combining traditional SE and LLMs in achieving reliable, efficient and effective LLM-based SE.& SLR on LLMs for SE.\\
				\hline
			\end{tabular}
	\caption{SLRs on AI for SE (sorted by publication year)} \label{figSLRAI4SE}
	\end{table}

Another important area of research is the development of roadmaps. Like SLRs, these seek to evaluate and categorise developments thus far. In addition, however, they project these developments into the future. The roadmaps presented in Table~\ref{figRoadmapsAI4SE} were recently published.

	\begin{table}[!ht]
			\centering 
			\footnotesize
			\rowcolors{1}{AntiqueWhite!30}{WhiteSmoke!30}
			\arrayrulecolor{Wheat}
			\begin{tabular}{|@{\hspace{2pt}}>{\raggedright\arraybackslash}p{0.1\linewidth}|@{\hspace{2pt}}>{\raggedright\arraybackslash}p{0.45\linewidth}|@{\hspace{2pt}}>{\raggedright\arraybackslash}p{0.45\linewidth}|}
				\hline
				\textbf{Paper} & \textbf{Summary} & \textbf{Difference to \textit{ai4se}} taxonomy\\
				\hline

				\citet{terragni_future_2025} & This paper presents a vision for an AI-driven software development framework. The main actors in this framework are software engineers, including developers, architects and testers, as well as a generic AI system, such as an LLM. & It discusses the challenges for requirements engineering, software testing, development and testing, and maintenance, which is more coarse-grained than \textit{ai4se}. \\
				\hline

				\citet{abrahao_software_2025} & This study examines how AI is reshaping the role of humans within the software ecosystem. It explores the challenges and opportunities arising from the interaction between technical and human factors in human-centred workflows.& This categorization of AI for SE focuses on how AI can enhance SE, render technologies and approaches obsolete, overturn common practice, retrieve past results, and change workflows. Human factors have yet to be incorporated into \textit{ai4se}, and could be considered a new dimension.\\
				\hline
				
				\citet{ahmed_artificial_2025} & This study highlights three key challenges for AI in SE: the use of GenAI and LLMs for engineering large software systems; the requirement for large, unbiased datasets and benchmarks for training and evaluating DL and LLMs for SE; and the need for a new code of digital ethics for applying AI in SE. & By analysing the challenges in classical and upcoming SE processes, the relevance of human factors in SE is also emphasised (see also~\citet{abrahao_software_2025}).\\
				\hline
				
				\citet{dirocco_use_2025} & This paper presents an overview of current LLM applications in MD[S]E and a roadmap for the deployment of LLMs to enhance the management, exploration, and evolution of modelling ecosystems.& The status of LLMs for MDSE is presented as a feature model. In particular, the more detailed modelling tasks could be considered new facets of MDSE in \textit{ai4se}. \\
				\hline
				
				\citet{marchezan_model-based_2024} & This paper explores the potential applications of GenAI in model-driven software maintenance and evolution.& As represented in \textit{ai4se}, this paper explores  how MDSE can be advanced through AI, particularly GenAI. It emphasises that AI can assist modellers, enhance their capabilities, and facilitate reasoning and automation in MDSE.\\	
				\hline
				
				\citet{bannon_infusing_2024} & This study analyses the infusion of AI techniques into DevOps and software security-related tasks. It also presents a call for action towards GenAI-based agents for SE. & The study emphasises the importance of human factors, security and trust considerations, all of which are candidate extensions for \textit{ai4se}.\\
				\hline
				
				\citet{lo_trustworthy_2023} & This paper reviews the development stages of AI for SE, highlighting trust and synergy as two key challenges that must be overcome before AI for SE tools can act as intelligent, responsible workmates. & The paper proposes research into trust-aware, privacy-aware, licence-aware, attack-resistant, secure and workflow-aware AI for SE solutions. This could potentially lead to new research being classified in the \textit{ai4se} taxonomy.\\
				\hline
				
				\citet{nguyen-duc_generative_2023} & With the help of an SLR, this paper identifies 78 open research questions across 11 SE areas for a GenAI-based SE solution. A research agenda is derived that highlights the need for solutions in the areas of dependability and accuracy, data accessibility and transparency, and sustainability aspects of GenAI solutions for AI. & These requirements could potentially lead to new research being classified in the \textit{ai4se} taxonomy. Moreover, the 'quality' aspects of AI for SE solutions, such as accuracy, robustness, transparency and sustainability, have the potential to be an extension of \textit{ai4se}.\\

				\hline
			\end{tabular}
	\caption{Research roadmaps on AI for SE (sorted by publication year)} \label{figRoadmapsAI4SE}
	\end{table}

Last but not least, some papers have investigated the development of a taxonomy for AI for SE, often focusing on a subfield of SE or the AI techniques used, as described in Table~\ref{figTaxonomyAI4SE}.

	\begin{table}[!ht]
			\centering 
			\footnotesize
			\rowcolors{1}{AntiqueWhite!30}{WhiteSmoke!30}
			\arrayrulecolor{Wheat}
			\begin{tabular}{|@{\hspace{2pt}}>{\raggedright\arraybackslash}p{0.1\linewidth}|@{\hspace{2pt}}>{\raggedright\arraybackslash}p{0.45\linewidth}|@{\hspace{2pt}}>{\raggedright\arraybackslash}p{0.45\linewidth}|}
				\hline
				\textbf{Paper} & \textbf{Summary} & \textbf{Difference to \textit{ai4se}} taxonomy\\
				\hline		
				
				\citet{treude_how_2025} & The types of interaction between software developers and AI are the focus of this taxonomy. & Human factors are a candidate extension for \textit{ai4se}.\\
				\hline
				
				\citet{melegati_dante_2024} & The DAnTE taxonomy focusses on the degree of automation of SE tools in general and defines six levels, including no automation, informer, suggester, and local, global and full generators. It also reviews AI for SE in particular.& DAnTE considers three types of generator, whereas \textit{ai4se} differentiates between partial and full generators only. The differences between 'global' and 'full' are subtle, and no tool currently addresses these levels.\\
				\hline
				
				\citet{zhao_natural_2022} & This taxonomy organizes 57 most frequently used NLP techniques in requirements engineering (RE). & This NLP classification could be used to further refine the AI types in \textit{ai4se}.\\
				\hline
				
				\citet{sofian_systematic_2022} & This systematic mapping study examines the use of four types of AI (and their combinations) in the development, deployment and maintenance of software.	& While the AI types and the SE activities in \textit{ai4se} are more detailed, this mapping's approach is similar. Still, for \textit{ai4se}, it could be considered to make the (AI) training a separate activity.\\
				\hline
				
				\citet{lopez-lujan_artificial_2021} & This systematic mapping study examines the use of AI in component-based SE and its potential to support this field. Not only does it map AI approaches to SE activities, it also identifies the problems being addressed.& \textit{ai4se} considers SE in general rather than component-based SE specifically. Nevertheless, some of the AI types and targets, which are the 'problems' in this study, may be considered for inclusion.\\
				\hline
				
				\citet{erlenhov_current_2019} & This study  focuses on bots that support software development (DevBots). It classifies contemporary DevBots using a facet-based taxonomy and outlines the ideal characteristics of future DevBots. The main facets are purpose, initiation, communication and intelligence. & While the main facets 'purpose' and 'intelligence' essentially correspond to the \textit{ai4se} dimensions 'purpose' and 'AI type', 'initiation' and 'communication' could be considered for inclusion in \textit{ai4se} as part of the human factors.\\
				\hline

				\citet{feldt_ways_2018} & This study introduces the AI-SEAL taxonomy, which classifies AI solutions for SE based on their application point, the type of AI technology employed, and the degree of automation they offer. & AI-SEAL uses three high-level SE processes, five AI tribes for categorising AI types and ten automation levels to determine the level of automation support. In contrast, \textit{ai4se} provides a detailed view of SE, categorising AI types into seven categories and automation levels into five.\\

				\hline
			\end{tabular}
	\caption{Taxonomies on AI for SE (sorted by publication year)} \label{figTaxonomyAI4SE}
	\end{table}

The SLRs, roadmaps, and taxonomies presented in Tables~\ref{figSLRAI4SE}, \ref{figRoadmapsAI4SE}, and \ref{figTaxonomyAI4SE} show that the intersection between AI and SE is a rapidly growing area of research and tool development. AI support in SE is used to address areas such as code generation, defect detection, code repair, code documentation, software testing, and challenges relating to software quality, security, and piracy. The application of AI in MDSE to provide guardrails for the application of AI by models is another focus.

Only a few taxonomies for SE and AI have been made explicit so far. They address a specific subfield of SE only or focus on classifying the AI methods and techniques used. Still, all of these taxonomies essentially align with \textit{ai4se}. In addition, the taxonomies in~\citet{feldt_ways_2018,melegati_dante_2024} and~\citet{erlenhov_current_2019} highlight the need to detail human factors when analysing the potentials of AI for SE. In future work, \textit{ai4se} is to be extended accordingly.
 
%%%%%%%%%%%%%%%%
\section{Summary and Outlook}\label{Sum}

This paper examined the intersection of big code, AI, SE and MDSE. To understand this evolving landscape, the novel \textit{ai4se} taxonomy has been introduced and applied to categorise recent literature. Furthermore, the concept of big models to highlight future opportunities for AI-augmented MDSE has been introduced. Future work will expand on this foundation by detailing additional SE activities within the taxonomy and categorizing more research results.

Integrating AI into SE shifts the software engineer's focus from routine coding to higher-level design and creative problem-solving, making the field more accessible and innovative. AI-augmented MDSE is poised to seamlessly bridge the gap between software models and code, with software model corpora facilitating the training of generative and agentic AI across the entire software lifecycle.

However, realizing this vision requires prioritizing human factors, which remain an open  challenge. The ethical implications of AI such as bias, fairness, transparency, and privacy cannot be treated as mere technical hurdles; they require human-centric solutions. Future research must deeply explore human-AI interaction paradigms, such as AI-assisted pair programming and pair modelling. By keeping the human in the loop, these collaborative approaches help maintain oversight, ensure software quality, and establish the accountability frameworks necessary to mitigate ethical risks.

To integrate trustworthy AI tools into everyday SE practice, the community must also overcome technical barriers, specifically regarding tool interoperability and standardization. Overcoming these hurdles will pave the way for AI-augmented software lifecycles that:
\begin{enumerate}
	\item automate repetitive and time-consuming tasks;
	\item leverage big code and big models for intelligent decision-making;
	\item detect requirements inconsistencies, design flaws, and defects early in the development process;
	\item manage the increasing complexity of modern systems through advanced analytics and monitoring; and
	\item enable continuous, iterative improvement of software processes via adaptive learning and feedback loops.
\end{enumerate}

To support the community in tracking the ongoing evolution of AI for SE and MDSE, the ontology and the classification of primary studies using the \textit{ai4se} taxonomy have been made publicly available under a CC-BY-SA licence for future use and extension.

\subsubsection*{Acknowledgement} The ideas presented in this paper have been developed through constructive dialogue in the Feldafinger Kreis, the German Testing Board and the Association for Software Quality and Education. The author would also like to express her sincere gratitude to the anonymous reviewers, whose insightful comments and constructive criticism significantly improved the quality and clarity of this work.

The author acknowledges that while authored by her, the writing process was aided by (AI) tools, specifically Google Scholar and ResearchRabbit for determining related work, and NotebookLM, ChatGPT and DeepL for fine-tuning the wording. 

The author has no competing interests to declare that are relevant to the content of this article.

%%%%%%%%%%%%%%%%
\bibliographystyle{sn-basic}
{\footnotesize
\bibliography{AI4SETaxonomy}
}

\end{document}